\documentclass[preprint,prx,floatfix]{revtex4-1}

\usepackage[utf8]{inputenc}
\usepackage{amsmath}
\usepackage{amssymb}
\usepackage{lipsum}
\usepackage{bm}
\usepackage{graphicx}
\usepackage{graphics}
\usepackage[dvipdf]{epsfig}
\usepackage{subcaption}
\usepackage{float}
\usepackage{wrapfig}
\usepackage{url}
\usepackage{epstopdf} 
\usepackage{xcolor}

\newcommand{\ddt}[2]{\frac{\mathrm{d^2}#1}{\mathrm{ d}#2^2}}
\newcommand{\ddx}[2]{\ddt{}x}
\newcommand{\ddy}[2]{\ddt{}y}
\newcommand{\ddz}[2]{\ddt{}z}

\begin{document}
\title{Spiral defect chaos in Rayleigh-B\'enard convection: \\Asymptotic and numerical studies of azimuthal flows induced by rotating spirals}

\author{Eduardo Vitral and Perry H. Leo}
\affiliation{Department of Aerospace Engineering and Mechanics,
    University of Minnesota, 110 Union St. SE, Minneapolis, MN 55455, USA}
\author{Jorge Vi\~nals}
\affiliation{School of Physics and Astronomy,
    University of Minnesota, 116 Church St. SE, Minneapolis, MN 55455, USA}
\author{Saikat Mukherjee}
\affiliation{Department of Biomedical Engineering and Mechanics,
    Virginia Tech, 495 Old Turner St., Blacksburg, VA 24061, USA}
\author{Mark R. Paul}
\affiliation{Department of Mechanical Engineering,
    Virginia Tech, 635 Prices Fork Road, Blacksburg, VA 24061, USA}
\author{Zhi-Feng Huang}
\affiliation{Department of Physics and Astronomy,
    Wayne State University, Detroit, Michigan 48201, USA}

\begin{abstract}
Rotating spiral patterns in Rayleigh-B\'enard convection are known to induce azimuthal flows, which raises the question of how different neighboring spirals interact with each other in spiral chaos, and the role of hydrodynamics in this regime. Far from the core, we show that spiral rotations lead to an azimuthal body force that is irrotational and of magnitude proportional to the topological index of the spiral and its angular frequency. The force, although irrotational, cannot be included in the pressure field as it would lead to a nonphysical, multivalued pressure. We calculate the asymptotic dependence of the resulting flow, and show that it leads to a logarithmic dependence of the azimuthal velocity on distance $r$ away from the spiral core in the limit of negligible damping coefficient. This solution dampens to approximately $1/r$ when accounting for no-slip boundary conditions for the convection cell's plate. This flow component can provide additional hydrodynamic interactions among spirals including those observed in spiral defect chaos. We show that the analytic prediction for the azimuthal velocity agrees with numerical results obtained from both two-dimensional generalized Swift-Hohenberg and three-dimensional Boussinesq models, and find that the velocity field is affected by the size and charges of neighboring spirals. Numerically, we identify a correlation between the appearance of spiral defect chaos and the balancing between the mean-flow advection and the diffusive dynamics related to roll unwinding.
\end{abstract}
\date{\today}

% \pacs{}

\maketitle

%%%%%%%%%%%%%%%%%%%%%%%%%%%%%%%%%%%%%%%%%%%%%%%%%%%%%%%%%%%%%%%%%%%%%%%%%%%

\section{Introduction}

An unexpected chaotic state near the onset of convection in a Rayleigh-B\'enard configuration was discovered in CO$_{2}$ gas (a low Prandtl number fluid) by Morris \textit{et al.} \cite{morris1993spiral}, in which rotating spirals are continuously nucleated and eliminated, yielding a state with persistent dynamics (i.e., \lq\lq spiral defect chaos''). Experimental evidence also suggests that, as the fluid Prandtl number decreases or the aspect ratio of the experimental cell increases, the chaotic state may emerge as the first bifurcation from the quiescent, conduction state \cite{re:liu96}, contrary to well established theory \cite{re:busse78,re:chandrasekhar81,re:cross93}. Classical stability theory is based on the Boussinesq model of thermal convection in a simple fluid \cite{re:chandrasekhar81}, and motion near the onset of Rayleigh-B\'enard convection is predicted to be variational, a fact that would preclude the observed persistent dynamics. As was recognized early on, the chaotic state is enabled through the coupling between the primary vertical velocity field mode that becomes unstable at threshold, and weakly damped, long wavelength rotational flows on the horizontal plane. That such near-marginal flows could be relevant in convection in large aspect ratio systems had been proposed earlier by Siggia and Zippelius \cite{siggia1981dynamics}.

Following the discovery of spiral defect chaos, a class of theoretical and computational analyses focused on two-dimensional (2D) models (the generalized Swift-Hohenberg models) that explicitly include the coupling between the vertical vorticity and an order parameter field appropriate for the convective instability (proportional to the vertical velocity or temperature deviation on the mid plane of the convection cell) \cite{swift1977hydrodynamic,manneville1983two,re:greenside85}. Numerical analysis of these models confirmed the importance of the coupling to vortical flows to model the transition to chaos \cite{xi1993spiral,re:schmitz02,cross1996theoretical,re:huang07,karimi2011exploring,karimi2019erratum}, although it remains unclear what its precise role actually is in sustaining the chaotic state. Extensive computational work also included direct numerical solution of the governing equations for a Boussinesq fluid in a B\'enard configuration \cite{re:decker94,re:schmitz02,chiam:2003,karimi2011exploring,*karimi2019erratum}, and also of the related problem of a single rotating spiral pattern filling the entire convection cell \cite{re:bestehorn92,re:bestehorn93,re:xi93,xi1993spiral}. In particular, a detailed numerical investigation by Karimi \textit{et al}. \cite{karimi2011exploring,*karimi2019erratum} has shown that the flow structure around a spiral core in a fully three-dimensional (3D) numerical solution of the Boussinesq equations is qualitatively similar to that of the simpler generalized Swift-Hohenberg models that incorporate 2D rotational flows. Yet, the main question as to the mechanisms underlying the appearance of the chaotic state and, in particular, the role of any hydrodynamic interactions among rotating spirals in an extended system remain unanswered.

In this work we present results based on both a generalized Swift-Hohenberg model in terms of a 2D order parameter field $\psi(\mathbf{x},t)$, which represents the vertical velocity of the fluid at the convection cell's mid-plane, and a full 3D solution of the Boussinesq equations. We begin by examining approximate solutions of the 2D model that correspond to a rotating spiral pattern, focusing on the rotational horizontal flow induced by an effective body force $\mathbf{f} = -(\nabla^2\psi){\bm \nabla}\psi$ (where $\nabla$ is the 2D gradient operator on the horizontal plane) that plays the role of the driving force of the vortical flow \cite{xi1993spiral,cross1996theoretical}. It is through this term that the curved convective rolls generate vertical vorticity, which in turn advects convective rolls. From the form of the $\psi$ field corresponding to a spiral pattern, we show that there exists an irrotational contribution to $\mathbf{f}$ that leads to a long-ranged azimuthal velocity field around the core of the spiral. For a laterally unbounded configuration, the azimuthal velocity decays as $v_{\varphi} \sim 1/r$ away from the core. If, instead, the azimuthal velocity is required to vanish at a finite distance $r_{b}$ from the core, we show that this asymptotic dependence is never attained for typical spiral sizes. Furthermore, if damping at the bottom and top bounding walls is neglected, we find $v_{\varphi} \sim r \ln (r/r_b)$, where $r_{b}$ is a cutoff distance at which the velocity vanishes. These results are verified numerically for both a rigidly rotating spiral pattern and the spiral defect chaotic state that are generated in the 2D generalized Swift-Hohenberg model and the 3D Boussinesq equations. 

In Sec. \ref{sec:gsh} we introduce the generalized Swift-Hohenberg model, which is the starting point of our asymptotic analysis. We expand the order parameter $\psi$ as a function of a small parameter in Sec. \ref{sec:ibf}, which allows us to express the force $\mathbf{f}$ in terms of gradients of the complex amplitude $A$ of the $\psi$ field. Based on a dynamic equation for the amplitude, we obtain the asymptotic form of $\mathbf{f}$ which exhibits both rotational and irrotational terms. We derive in Sec. \ref{sec:vel} different components of the azimuthal velocity from the asymptotic form of $\mathbf{f}$, and show that they are dominated by the contribution from the irrotational force term. The numerical methods used for both generalized Swift-Hohenberg and Boussinesq models are detailed in Secs. \ref{sec:gsh2} and \ref{sec:bsq}, respectively. In Sec. \ref{sec:azi}, we confirm the analytic results for the dependence of the azimuthal velocity $v_{\varphi}$ on damping by computing it  for a range of damping coefficients. While for relatively large damping the azimuthal flows are largely confined near the core of each spiral, as damping decreases a hydrodynamic interaction between neighboring spirals arises through the cutoff length $r_{b}$.  In Sec. \ref{sec:adv}, we discuss the role of advection versus spiral arm unwinding in the rotation of the spiral. Finally, in Sec. \ref{sec:bsq2} we present a comparison of the azimuthal velocity computed from the generalized Swift-Hohenberg model and the full Boussinesq equations.

\section{Azimuthal flow induced by a rotating spiral}

\subsection{The generalized Swift-Hohenberg model}
\label{sec:gsh}

The Swift-Hohenberg model for Rayleigh-B\'enard convection \cite{swift1977hydrodynamic} follows from a 2D projection of the governing fluid equations in the Boussinesq approximation that eliminates the dependence of the temperature, pressure, and velocity fields on the vertical coordinate $z$ near the onset of convection. This results in a gradient model for an order parameter field $\psi(\mathbf{x},t)$ [with $\mathbf{x} = (x,y)$] that represents the vertical velocity on the mid-plane of the convection cell \cite{manneville1983two,manneville1984modelisation,manneville1990dissipative}. The model was later generalized to account for the coupling between the unstable mode at the onset of convection and 2D mean flows \cite{manneville1983two,re:greenside85}, and is associated with the equations
\begin{eqnarray}
  \label{eq:sh}
  \partial_t\psi + \mathbf{v}\cdot{\bm \nabla}\psi &=& \epsilon\psi -(\nabla^2+q_0^2)^2\psi - \psi^{3},
  \\
  \label{eq:vort}
  \Big[\partial_t - \sigma(\nabla^2-c^2) \Big]\nabla^2\zeta &=& g_m \Big[{\bm \nabla} (\nabla^2\psi)\times{\bm \nabla}\psi\Big]\cdot\mathbf{\hat{z}},
\end{eqnarray}
where $\epsilon$ is a bifurcation parameter that measures the dimensionless distance to the convection threshold (in terms of the Rayleigh number), $\mathbf{v}(\mathbf{x},t)$ is the 2D incompressible mean flow velocity, and $\sigma$ is a rescaled Prandtl number. The mean flow velocity is obtained from the vertical vorticity potential $\zeta(\mathbf{x},t)$ via $\mathbf{v} = \bm{\nabla}\times(\zeta \mathbf{\hat{z}})$, such that the vertical vorticity $\Omega_z = ({\bm \nabla} \times \mathbf{v}) \cdot \hat{\mathbf z} = -\nabla^2 \zeta$. A momentum damping coefficient $c^{2}$ is introduced to model viscous friction at the top and bottom bounding walls, and appears from averaging derivatives of the flow in the vertical direction over the thickness of the convection cell.  In the case of free-slip (i.e., stress-free) boundary conditions, one would have $c^2=0$, while $c^2 > 0$ for no-slip boundary conditions. The coefficient $g_{m}$ controls the magnitude of the flow coupling \cite{re:greenside85,manneville1984modelisation}, which increases as the Prandtl number decreases and also appears from the averaging process.

The right hand side of Eq. (\ref{eq:vort}) can be written as $-g_m({\bm \nabla} \times \mathbf{f}) \cdot \hat{\mathbf z}$, with $\mathbf{f} = -(\nabla^2\psi){\bm \nabla}\psi$. This effective body force originates from projecting the advection nonlinearity in the Boussinesq model onto the 2D order parameter model. This force has a functional analog in models of active matter in which an active stress breaks equilibrium symmetry relations \cite{marchetti2013hydrodynamics,tiribocchi2015active,kirkpatrick2019driven} and hence directly allows non-variational flows.

\subsection{Effective body force induced by a rotating spiral}
\label{sec:ibf}

Rotating spiral and target solutions are well known to emerge from Eq.~(\ref{eq:sh}) \cite{pismen1999vortices}. Away from the core, the solution for the order parameter field in polar coordinates $(r,\varphi)$ has the form $\psi = A e^{iq_0r} + \textrm{c.c.}$, where $A$ is a slowly-varying complex amplitude that can be written as $A(r,\varphi,t) = \rho(r)e^{i\theta}$, with phase $\theta = m\varphi - \omega t$ and a real amplitude $\rho$. Here $\omega$ is the angular frequency of the spiral, and the topological charge $m$ is an integer representing the index of the singularity \cite{pismen1991mobility} (which is the number of arms in the spiral in this case). While target patterns with $m = 0$ present a single-valued $\theta$, rotating spirals have a multivalued phase $\theta$ as $m \neq 0$. This has implications for their topological stability \cite{pismen1991mobility} as the circulation of $\theta$ around a contour $\gamma$ enclosing the spiral core has a quantized value that depends on the topological charge, i.e., $\oint_\gamma \bm{\nabla}\theta \cdot d\mathbf{l} = 2 \pi m$, where $\mathbf{l}$ is the vector function that defines the path. Velocity fields induced by rotating spirals have been argued to decay with distance as $1/r$, and to be negligible for spiral rotation as compared to motion induced by wavevector frustration \cite{cross1996theoretical}. This implies that direct hydrodynamic interactions among rotating spirals are negligible, and therefore the role of mean flows in inducing and sustaining the chaotic state remains to be understood. We reexamine this issue by investigating the azimuthal velocity generated by a rotating spiral by both asymptotic and numeric analyses.

We derive the asymptotic form of the body force $\mathbf{f}$ that appears in the 2D momentum Eq.~(\ref{eq:vort}). Near the convection threshold $\epsilon \ll 1$, we expand the order parameter $\psi$ into a periodic base state in terms of a slowly varying amplitude of the form
\begin{equation}
 \psi = \epsilon \left[ A(\mathbf{X},T) e^{i\mathbf{k}\cdot\mathbf{x}} + {\rm c.c.}\right ],
\label{eq:expansion}
\end{equation}
where ($\mathbf{X}$,T) denotes slow spatial and time scales upon which the amplitude $A$ depends. Assuming an isotropic expansion in which ${\bm \nabla} \rightarrow  \pm i \mathbf{k} + \epsilon \bm{\nabla}$, $| \mathbf{k} | = q_{0}$, and the gradient only acts on the slow $\mathbf{X}$ scale, we find
\begin{eqnarray*}
\nabla^{2} \left[ \epsilon A(\mathbf{X},T) e^{i \mathbf{k} \cdot\mathbf{x}} \right] &=& - \epsilon k^{2} A e^{i \mathbf{k} \cdot\mathbf{x}} + 2i \epsilon^{2} \mathbf{k}\cdot{\bm \nabla} A e^{i\mathbf{k} \cdot\mathbf{x}} + \epsilon^3 \nabla^2 A e^{i\mathbf{k} \cdot\mathbf{x}},
\end{eqnarray*}
which leads to the following resonant terms in the amplitude expansion of the force $\mathbf{f}$ (those originated from a combination of wavevectors whose result is zero)
\begin{eqnarray*}
\nabla^{2} [\epsilon A(\mathbf{X},T) e^{i \mathbf{k} \cdot\mathbf{x}} ] {\bm \nabla} [\epsilon A^*(\mathbf{X},T) e^{-i \mathbf{k} \cdot\mathbf{x}} ] & = & i \epsilon^{2}k^{2} |A|^{2}\,{\bf k} - \epsilon^{3}k^{2} A\bm{\nabla} A^{*} + 2\epsilon^3({\bf k}\cdot\bm{\nabla} A) \, A^*{\bf k} 
\\
&&+ 2i \epsilon^{4} ({\bf k} \cdot \bm{\nabla} A) \bm{\nabla} A^{*} - i \epsilon^{4}  A^{*} \nabla^{2} A \, {\bf k} + \epsilon^{5} \bm{\nabla}^{2}A \bm{\nabla} A^{*}.
\end{eqnarray*}
Explicitly adding the complex conjugate terms, we find
\begin{eqnarray*}
  (\nabla^{2} \psi ) \bm{\nabla} \psi
  &=&
      - \epsilon^{3}k^{2} \bm{\nabla} |A|^{2}
      + 2\epsilon^3\Big[({\bf k}\cdot \bm{\nabla} A)A^*+({\bf k}\cdot \bm{\nabla} A^*)A\Big]\,{\bf k}
  \\
  &&
      + 2 i \epsilon^4\Big[({\bf k}\cdot \bm{\nabla} A)\bm{\nabla} A^*-({\bf k}\cdot \bm{\nabla} A^*)\bm{\nabla} A\Big]
     + i \epsilon^{4} \left( A \nabla^{2} A^{*} - A^{*} \nabla^{2} A \right) {\bf k}
  \\
  &&
      + \epsilon^{5} \left( \nabla^{2}A \bm{\nabla} A^{*} + {\rm c.c.} \right).
\end{eqnarray*}
This expression can be further simplified by noting that in the radial direction the rigid rotating spiral is approximately a solution of the governing equation for the order parameter. That is, away from the spiral's core we have ${\bf k} = q_0 \hat{\bm{r}}$ and $A = \rho(r)e^{i\theta}$, where $\theta = m\varphi - \omega t $. This leads to 
 $ ( A \nabla^{2} A^{*} - A^{*} \nabla^{2} A ) = \bm{\nabla}\cdot(A\bm{\nabla} A^* - A^*\bm{\nabla} A) = 0$.
By gathering terms up to order $\epsilon^5$ and rescaling all the quantities back to the original scales ($\mathbf{x}$,t), we find that the force can be written as
\begin{eqnarray}
\nonumber
  \mathbf{f}
  &=& - (\nabla^{2} \psi) \bm{\nabla} \psi 
  \\
  &=&
      q_0^{2} \bm{\nabla} (|A|^{2} - 2 \rho^2)
     - 4 q_0 \frac{m \rho\,\rho'}{r} \hat{\bm{\varphi}}
     - \left( \nabla^{2}A \bm{\nabla} A^{*} + {\rm c.c.} \right).
     \label{eq:sh_f}
\end{eqnarray}
The first term in the RHS is of gradient form and does not contribute to Eq. (\ref{eq:vort}). In a pressure-velocity formulation, it can be absorbed into the pressure term. The other two terms contribute to the mean flow. 

We begin with the last term of Eq. (\ref{eq:sh_f}), which can be written as a function of the angular frequency of the spirals. From Eq. (\ref{eq:sh}), the corresponding amplitude equation \cite{manneville1990dissipative,hoyle2006pattern} can be written in polar coordinates for curved rolls (targets or spirals) \cite{vitral2019role,korzinov1993origin},
\begin{equation}
\partial_{t} A = \epsilon A + 4q_0^2(\partial_r^2+r^{-1}\partial_r)A - 2i q_0 r^{-2} (2 \partial_r + r^{-1})\partial_\varphi^2A - r^{-4}\partial_\varphi^4A-3 |A|^{2} A.
\label{eq:amp}
\end{equation}
This amplitude equation appears from a solvability condition at $\mathcal{O}(\epsilon^{3/2})$ of the expansion, with $\psi$ expanded in power of $\epsilon$ similarly to Eq. (\ref{eq:expansion}). Rearranging terms, we obtain the following expressions for a rigidly rotating spiral with $\partial_{t} A = - i \omega A$,
\begin{eqnarray}
\nonumber
(\partial_r^2+r^{-1}\partial_r)A &=& \frac{1}{4 q_0^2}\bigg[-(i \omega + \epsilon)A 
+2i q_0 r^2(2\partial_r+r^{-1})\partial^2_\varphi A \\ %\nonumber
&& \hspace{9mm} + r^{-4}\partial^4_\varphi A + 3|A|^{2} A \bigg].
%\\ \nonumber
%(\partial_r^2+r^{-1}\partial_r)A^{*} &=& \frac{1}{4 q_0^2} %\bigg[ (i \omega - \epsilon) A^{*} 
%+2i q_0 r^2(2\partial_r+r^{-1})\partial^2_\varphi A^* \\
%&& \hspace{9mm} + r^{-4}\partial^4_\varphi A^* + 3|A|^{2} %A^{*}\bigg].
\label{eq:polard}
\end{eqnarray}
For the complex amplitude of a spiral given by $A = \rho(r)e^{i\theta}$, we have $\bm{\nabla} A = (\rho^\prime\hat{\bm{r}} + i m r^{-1}\rho \hat{\bm{\varphi}})e^{i\theta}$ and $|\bm{\nabla}\theta| = m/r$. Hence, from Eqs. (\ref{eq:sh_f}) and (\ref{eq:polard}) we obtain
\begin{equation}
\mathbf{f} = - \frac{1}{2 q_0^2}  \left[ \left( 3\rho^{2} - \epsilon + |\nabla\theta|^4 - 4q_0^2|\nabla\theta|^2\right)\rho\rho^{\prime} \hat{\bm{r}} + \left(-\frac{\rho^2  m \omega}{r} + \frac{8q_0^3 m \rho\rho^\prime}{r}\right) \hat{\bm{\varphi}} \right].
\label{eq:finalf}
\end{equation}
This is the central result of this section. The radial component in the RHS of Eq. (\ref{eq:finalf}) is irrotational and can be included in a redefinition of the pressure. The azimuthal component vanishes for targets with $m = 0$ (i.e., no angular dependence). It also vanishes near the core ($r \rightarrow 0$) since $\rho \rightarrow 0$ as the core is approached. However, away from the core where $\rho$ is approximately constant, the term [$-(\rho^2 m \omega/r) \hat{\bm \varphi}$] can be written as $-\rho^{2} m \omega \nabla \varphi$. This is an azimuthal body force induced by the rotating spiral that is irrotational. This irrotational force term cannot be eliminated by subsuming it into the pressure as the latter would become multivalued. That is, the observed pressure is continuous, without a direct dependence on $\varphi$, which would lead to a jump of $2\pi$ (see Fig. \ref{fig:psi-theta}). The curl of this irrotational force corresponds to a vorticity point source at the origin. No true divergence exists in this term as the amplitude $\rho$ vanishes at the core. We will retain this irrotational force, and calculate its contribution to the azimuthal velocity explicitly.

\subsection{Azimuthal velocity field}
\label{sec:vel}

In order to compute the velocity field that results from the force given in Eq.~(\ref{eq:finalf}), we first obtain an asymptotic expression for the amplitude $\rho$ by substituting $A = \rho(r)e^{i\theta}$ in the amplitude equation (\ref{eq:amp}); in the stationary limit we find
\begin{equation}
 4 q_0^2(\partial_r^2+r^{-1}\partial_r)\rho+(\epsilon-|\nabla\theta|^4-3\rho^2)\rho  =  0.
\end{equation}
For $r \gg 1$, using $\theta = m \varphi - \omega t$ we obtain
\begin{eqnarray}
\rho^2 &=& \frac{1}{3}\epsilon - \frac{1}{3}|\bm{\nabla}\theta|^4 = \frac{4}{3}\epsilon - \frac{4 m^4}{3 r^4}, \quad 2\rho\rho^\prime =\frac{4 m^4}{3 r^5}.
\label{eq:rho2}    
\end{eqnarray}
Substituting Eq. (\ref{eq:rho2}) into Eq. (\ref{eq:finalf}) yields 
\begin{equation}
f_\varphi  = \frac{1}{2 q_0^2}\bigg(\frac{\rho^2\ m\omega}{r}-\frac{8q_0^3m\rho\rho^\prime}{r}\bigg) = \frac{1}{6 q_0^2}\left[\frac{\epsilon m\omega}{r}-\frac{ m^5\omega}{r^5} - \frac{16 q_0^3 m^5}{r^6}\right].
\label{eq:fphi}
\end{equation}
The radial component of $\mathbf{f}$ given in Eq. (\ref{eq:finalf}) can be written in a gradient form and absorbed into the pressure term; thus we only need to consider $\mathbf{f} = f_\varphi\hat{{\bm{\varphi}}}$. As shown in Eq.~(\ref{eq:fphi}), this azimuthal force consists of an irrotational contribution (the first term) and two rotational contributions.

The calculation of the azimuthal velocity is simplified by using a pressure-velocity representation of Eq.~(\ref{eq:vort}) for Stokes flow,
\begin{eqnarray}
-\bm{\nabla} p + \sigma(\nabla^2 -c^2)\mathbf{v} + g_{m} \mathbf{f} = 0. \label{eq:stokes}
\end{eqnarray}
Since this equation is linear in $\mathbf{v}$ we solve separately for the three components of $f_\varphi$ in Eq.~(\ref{eq:fphi}), which leads to the three velocity contributions $\mathbf{v}_{1}$, $\mathbf{v}_{2}$, and $\mathbf{v}_{3}$. The component $\mathbf{v}_{1}$ satisfies
\begin{equation}
-\bm{\nabla} p + \sigma(\nabla^2 -c^2)\mathbf{v}_1 + \frac{\epsilon g_m m\omega}{6 q_0^2 r} \hat{\bm{\varphi}} =0.
\label{eq:stkr1}
\end{equation}
This flow component is induced by the irrotational part of the azimuthal force. In the vorticity and stream function formulation of Eq.~(\ref{eq:vort}), the corresponding term is zero except for a point source of vorticity at the origin. In this configuration, the pressure changes only along the radial direction (as observed in Fig.~\ref{fig:pressure}), so that the azimuthal component of the velocity satisfies an inhomogeneous modified Bessel equation
\begin{equation}
\partial^2_r v_{1\varphi} + \frac{1}{r} \partial_r v_{1\varphi} - (c^2+\frac{1}{r^2})v_{1\varphi} = -\frac{\epsilon g_m m\omega}{6 q_0^2\sigma r}.
\end{equation}
Assuming Dirichlet boundary conditions, so that the velocity approaches zero at the spiral's core $r = 0$ and vanishes at some distance $r_b$, we find
\begin{eqnarray}
    v_{1\varphi} &=& \frac{m\omega \epsilon g_m}{6 q_0^2 \sigma c^2}
    \Bigg[ \frac{1}{r}+\Bigg(c K_1(r_b c) - \frac{1}{r_b}\Bigg)\frac{I_1(c\,r)}{I_1(r_b c)} - c K_1(c\,r) \Bigg],
\label{eq:bessel}
\end{eqnarray}
where $I_1$ and $K_1$ are modified Bessel functions of first and second type, respectively. In the limit of $c^2 \rightarrow 0$, for which damping at the top and bottom bounding walls is negligible (free-slip), the solution for $v_{1\varphi}$ reduces to \cite{re:misc1}
\begin{equation}
v_{1\varphi} = -\frac{m\omega \epsilon g_m}{12 q_0^2 \sigma} r \ln (r/r_b) .
\label{eq:rlogr}
\end{equation}
For $c^{2} > 0$, in the limit $r_{b} \rightarrow \infty$ the contributions from the parts containing the modified Bessel functions in Eq. (\ref{eq:bessel}) become negligible at long distance, so that $v_{1\varphi} \sim 1/r$, in agreement with the result in Ref.  \cite{cross1996theoretical}. Recall that several approximations made here hold only away from the spiral core, and therefore this solution must be regarded as an outer solution for the flow.

It is possible to obtain analytically a solution for the rotational component of the flow $v_{2\varphi}$, although only when $c^{2} = 0$. The corresponding flow equation is given by
\begin{equation}
-\bm{\nabla} p + \sigma(\nabla^2 -c^2)\mathbf{v}_2 - \frac{g_m m^5\omega}{6 q_0^2 r^5}\hat{\bm{\varphi}} = 0.
\end{equation}
It can be rewritten in terms of $\zeta$ in polar coordinates, i.e.,
\begin{equation}
\partial^4_r \zeta+\frac{2}{r}\partial^3_r\zeta
-\frac{1}{r^2}\partial^2_r\zeta + \frac{1}{r^3}\partial_r\zeta
+\frac{1}{r^3}\partial_r\partial_\varphi^2\zeta+\frac{1}{r^2}\partial^2_r\partial^2_\varphi\zeta
-c^2 \bigg(\frac{1}{r}\partial_r\zeta+\partial^2_r\zeta \bigg)
= \frac{2 g_m m^5\omega}{3q_0^2 \sigma r^6},
\end{equation}
where $\zeta = \zeta(r)$ due to $\mathbf{v}_2 = v_{2\varphi}\hat{\bm{\varphi}}$. At large distances and $c^{2} = 0$, we find
\begin{equation}
    \zeta(r) = \frac{g_m m^5\omega}{96 q_0^2 \sigma r^2}.
\end{equation}
Therefore, since $v_{2\varphi} = -\partial_r \zeta$ we obtain
\begin{equation}
    v_{2\varphi} = \frac{g_m m^5\omega}{48 q_0^2 \sigma r^3}.
    \label{eq:rm3}
\end{equation}

Similar to $v_{2\varphi}$, we are able to obtain a solution for the other rotational component $v_{3\varphi}$ generated by the last term in Eq. (\ref{eq:fphi}), also for $c^{2} = 0$. Following the same steps, we obtain
\begin{equation}
    v_{3\varphi} = \frac{8 g_m m^5q_0}{45 \sigma r^4}.
    \label{eq:rm4}
\end{equation}

In summary, the azimuthal flow induced by a rotating spiral can be decomposed into two separate contributions arising from irrotational and rotational force components respectively. The former, $v_{1\varphi}$ as given by Eq. (\ref{eq:bessel}), leads to a long ranged logarithmic dependence of the azimuthal velocity when $c^{2} = 0$, and to a $1/r$ decay at finite damping with $c^{2} > 0$ as $r_{b} \rightarrow \infty$. Rotational forces lead to azimuthal velocities $v_{2\varphi}$ and $v_{3\varphi}$ that decay as power laws ($1/r^{3}$ and $1/r^{4}$) for $c^2 = 0$, as shown in Eqs. (\ref{eq:rm3}) and (\ref{eq:rm4}), and therefore decay much faster than the flow $v_{1\varphi}$ created by the irrotational component of the force. We will use these results to interpret the numerical calculations in the next section.

\section{Numerical methods}
\label{sec:num}

\subsection{2D generalized Swift-Hohenberg model}
\label{sec:gsh2}

\begin{figure}[ht]
	\centering
    \begin{subfigure}[b]{0.33\textwidth}
    \includegraphics[width=\textwidth]{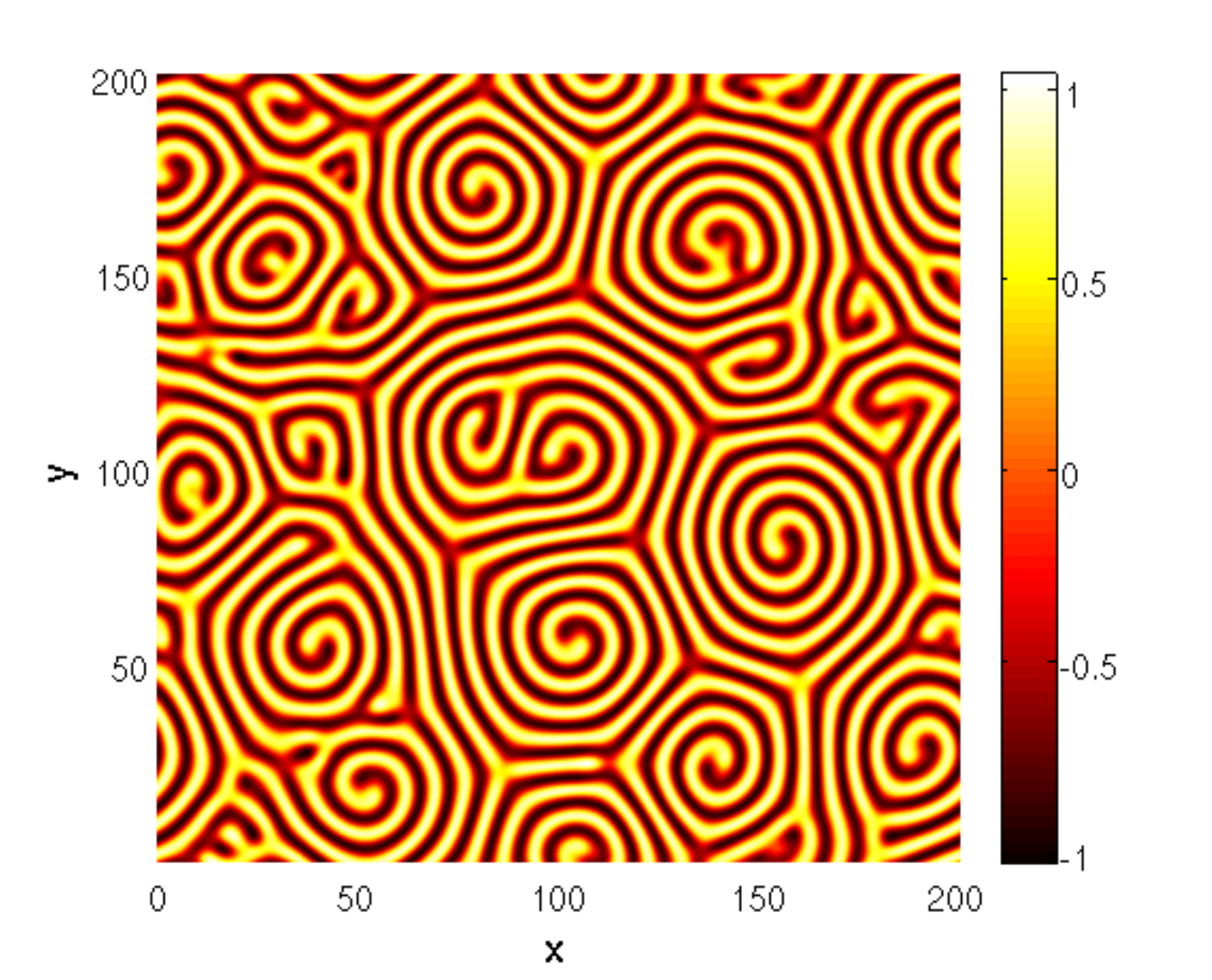}
    \caption{$\psi$}
    \label{fig:psi}
    \end{subfigure}
    \hspace{-1mm}
    \begin{subfigure}[b]{0.31\textwidth}
    \includegraphics[width=\textwidth]{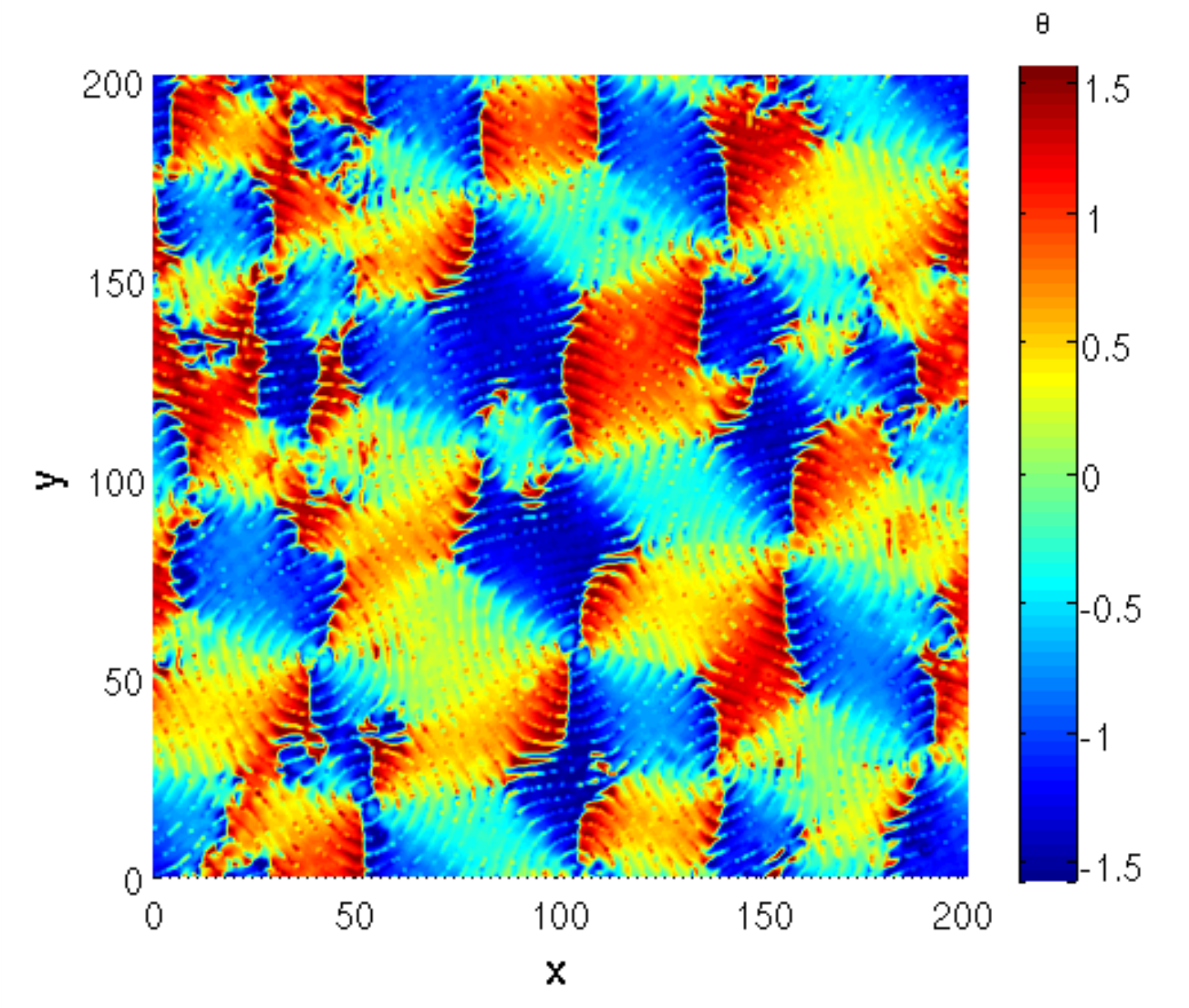}
    \caption{$\theta$}
    \label{fig:theta}
    \end{subfigure}
    \hspace{-1mm}
    \begin{subfigure}[b]{0.32\textwidth}
    \includegraphics[width=\textwidth]{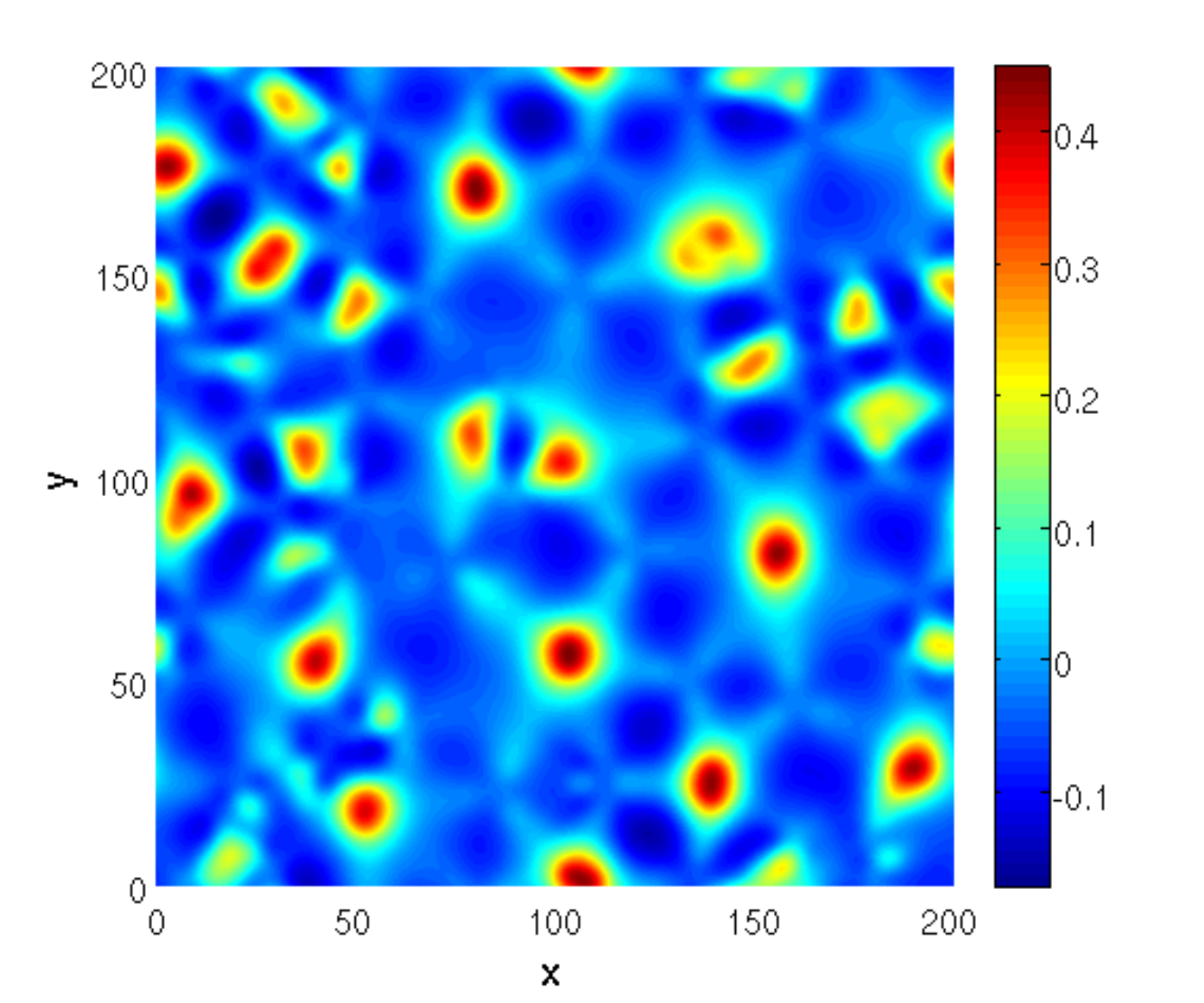}
    \caption{Pressure}
    \label{fig:pressure}
    \end{subfigure}
    \caption{Spatial pattern of (a) the order parameter field $\psi$ comprising several one-armed spirals, (b) the corresponding local phase $\theta$, and (c) the pressure. The model parameters used are $\epsilon = 0.7$, $g_m = 50$, $c^2 = 2$, and $\sigma = 1$.}
	\label{fig:psi-theta}
\end{figure}

Computations of the generalized Swift-Hohenberg model were based on the vorticity formulation, Eqs. (\ref{eq:sh}) and (\ref{eq:vort}). We also conducted spot checks with an equivalent 2D pressure-velocity formulation based on Eqs. (\ref{eq:sh}) and (\ref{eq:stokes}), and computed the effective pressure field as shown in Fig. \ref{fig:pressure}. The results obtained for the velocity field are identical within numerical accuracy. For all the results presented, the equations have been solved on an equally spaced, square grid of $512^2$ nodes, with $q_0 = 1$ and the grid spacing $\Delta x = 2 \pi/16$. We used a pseudo-spectral method, where gradient terms are computed in Fourier space with a second-order implicit iteration scheme, and nonlinearities are computed in real space through an explicit second-order Adams-Bashforth scheme. The time step used is $\Delta t = 10^{-3}$. The algorithm was implemented by using the parallel FFTW routine with associated MPI libraries. Periodic boundary conditions were used throughout. In our calculations the parameters were chosen as $g_m = 50$, $\sigma = 1$, and $\epsilon = 0.7$. Further details about the influence of the various parameters on the qualitative nature of the patterns obtained have been given in Refs. \cite{cross1996theoretical,karimi2011exploring}. From the pressure-velocity formulation, the pressure has been computed through the pressure Poisson equation which follows from the longitudinal projection of the underlying momentum conservation equation (by taking the divergence of Eq. (\ref{eq:stokes}) and accounting for incompressibility $\bm{\nabla}\cdot \mathbf{v}=0$). The same grid setup, model parameters, and boundary conditions were used in this case.

Figure \ref{fig:psi} shows a typical configuration of the $\psi$ field inside the regime of spiral defect chaos. It is obtained by time integration of the model equations from a random initial condition of uniformly distributed $\psi \in (-0.05,0.05)$ and zero initial velocity. The figure shows multiple one-armed spirals, obtained at time $t= 10^5$ for $c^{2} = 2$. Following the algorithm of Egolf \textit{et al.} \cite{egolf1998importance}, we show in Fig. \ref{fig:theta} the corresponding spatial distribution of phase $\theta$, and observe the expected discontinuity of $2\pi$ when enclosing a full circle around the core of each spiral. Although the phase is multivalued, the body force is continuous and the resulting pressure (illustrated in Fig. \ref{fig:pressure}) is also continuous. Note that the pressure is mostly radially symmetric, with its local maximum near the core of every spiral.

\begin{figure}[ht]
    \begin{subfigure}{0.45\textwidth}
        \centering
        \includegraphics[width=0.9\linewidth]{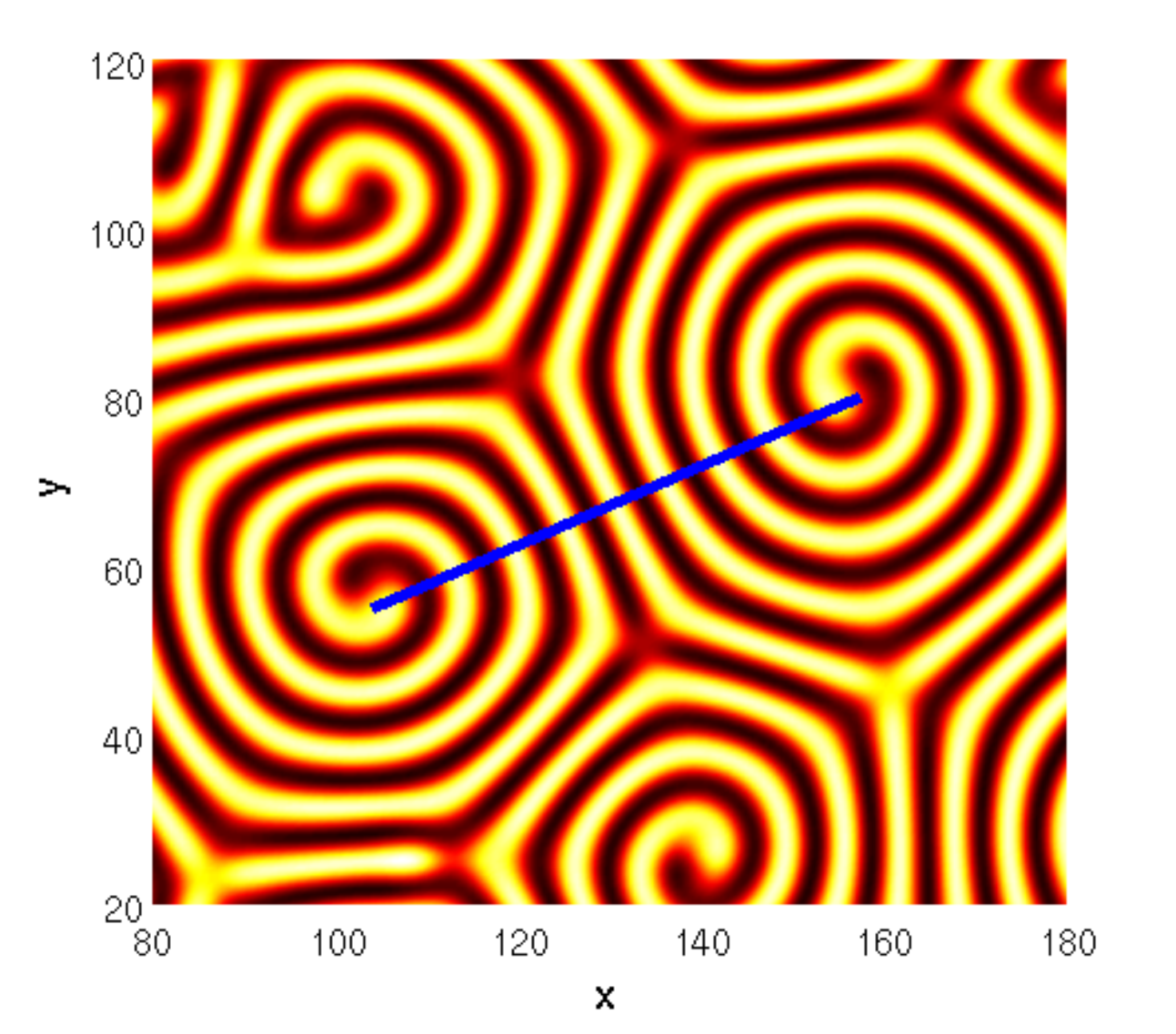}
    \end{subfigure}
    \begin{subfigure}{0.45\textwidth}
        \centering
        \includegraphics[width=0.9\linewidth]{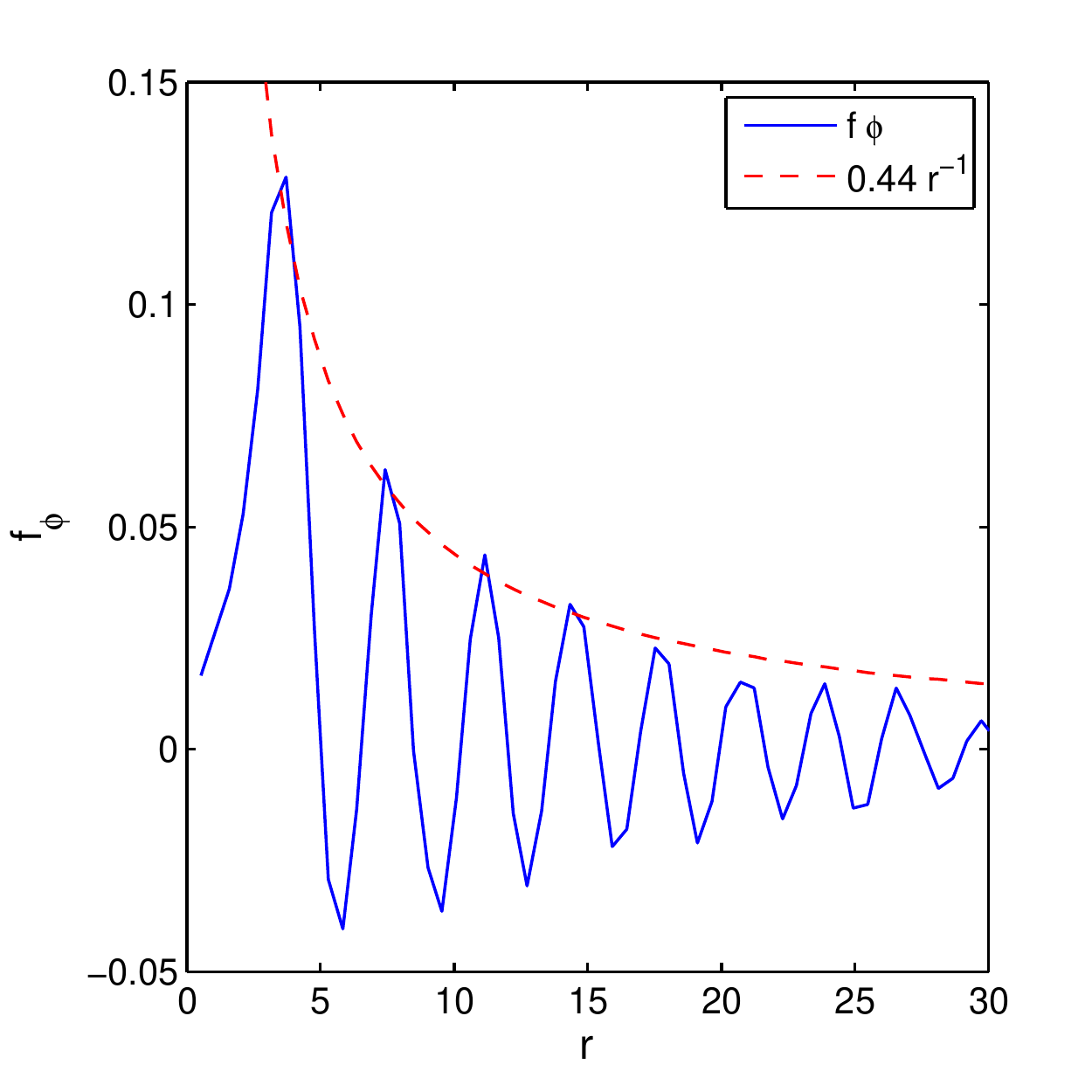}
    \end{subfigure}
    \caption{Left: Two clockwise rotating spirals with cores located at approximately $(104,56)$ and $(156,81)$. The blue line has a length of 58, which is roughly the distance between the cores. Right: Azimuthal component of the force $\mathbf{f}=-\nabla^2\psi\bm{\nabla}\psi$ (solid line), with $r = 0$ at the $(104,56)$ core. The dashed line is a guide to the eye showing the approximate $0.44/r$ decay of the force amplitude.}
    \label{fig:same}
\end{figure}

From each spatial configuration, such as the one shown in Fig. \ref{fig:psi}, we extract the locations of spiral cores by plotting the magnitude of the velocity $\mathbf{v}$ and searching for the location inside the vortices where $|\mathbf{v}| = 0$. The flow generated by each spiral has the form of a vortex \cite{karimi2011exploring,karimi2019erratum}; hence well-formed spirals are detectable through axially symmetric rings in $|\mathbf{v}|$ or smooth spikes in the vorticity potential (for which the core is located at the maximum of $\zeta$).

Figure \ref{fig:same} shows two neighboring spirals rotating in the same direction, with cores located at positions $(104,56)$ and $(156,81)$ of Fig. \ref{fig:psi}. By setting the origin, $r = 0$, of a polar coordinate system at the core of the left spiral at $(104,56)$, the figure also shows the radial dependence of the azimuthal component of the force $\mathbf{f}$ up to the edge of the spiral. The amplitude of the force decays slowly with distance $r$, and for $r > 5$ oscillates with periodic wavelength slightly larger than $\pi / q_{0}$, half of the approximate stripe/roll periodicity $\lambda_0$ from the linear solution, as expected from $\mathbf{f}=-\nabla^2\psi\bm{\nabla}\psi$. The velocity generated by this rotating spiral will be investigated in Sec. \ref{sec:vel}.

\subsection{3D Boussinesq equations}
\label{sec:bsq}

Rayleigh-B\'enard convection is the buoyancy driven convection that occurs when a shallow and horizontal layer of fluid is heated from below. The fluid motion is described by the Boussinesq equations~\cite{re:cross93}, which represent the conservation of momentum, energy, and mass, and are given as 
\begin{eqnarray}
\text{Pr}^{-1} \! \left( \frac{\partial \mathbf u}{\partial t} \!+\! \mathbf u \cdot \bm{\nabla} \mathbf u \right )\! &=&\! - \bm{\nabla} p \!+\! \nabla^2 \mathbf u \!+\! \text{Ra} T \hat{\mathbf{z}}, \label{eq:momentum} \\
\frac{\partial T}{\partial t} + \mathbf{u} \cdot \bm{\nabla} T &=& \nabla^2 T, \label{eq:energy} \\
\bm{\nabla} \cdot \mathbf{u} &=& 0. \label{eq:mass}
\end{eqnarray}
In these equations, $\mathbf{u}(x,y,z,t) = (u,v,w)$ is the velocity vector with components $(u,v,w)$ in the $(x,y,z)$ directions, respectively. The pressure is given by $p(x,y,z,t)$, the temperature field is denoted by $T(x,y,z,t)$, and $\hat{\mathbf{z}}$ is a unit vector in the positive $z$ direction which opposes the direction of gravity. Equations~(\ref{eq:momentum})--(\ref{eq:mass}) have been nondimensionalized using the depth of the convection layer $d$ as the length scale and the vertical heat diffusion time $d^2/ \kappa$ as the time scale, where $\kappa$ is the thermal diffusivity.  The vertical diffusion time represents the time required for heat to diffuse from the bottom to the top of the convection layer. Additionally, the constant temperature difference between the bottom and top boundaries, $\Delta T$, is set as the temperature scale. Using this convention, $0 \leq T \leq 1$ where $T(z\!=\!0)\!=\!1$ at the bottom boundary and $T(z\!=\!1)\!=\!0$ at the top boundary. 
The Rayleigh number $\text{Ra} \!=\!\beta g d^3 \Delta T/(\nu \kappa)$ is often the control parameter used in experiments and represents the ratio of buoyancy to thermal and viscous dissipation, where $\beta$ is the thermal expansion coefficient and $\nu$ is the kinematic viscosity. It is often convenient to use the reduced Rayleigh number $\epsilon = (\text{Ra}-\text{Ra}_c)/\text{Ra}_c$ to describe the degree of driving beyond the convective threshold, where $\text{Ra}_c$ is the critical Rayleigh number. The way this number rescales to $\epsilon$ in Eq. (\ref{eq:sh}) is detailed in Appendix \ref{sec:ap}. For an infinite layer of fluid with no-slip boundaries $\text{Ra}_c=1707.76$ and the nondimensional critical wave number of the convection rolls is $q_c=3.1165$~\cite{re:cross93}.  Therefore, the width of a single convection roll will be approximately unity after the nondimensionalization.

\begin{figure}[ht]
\begin{center}
\includegraphics[width=5in]{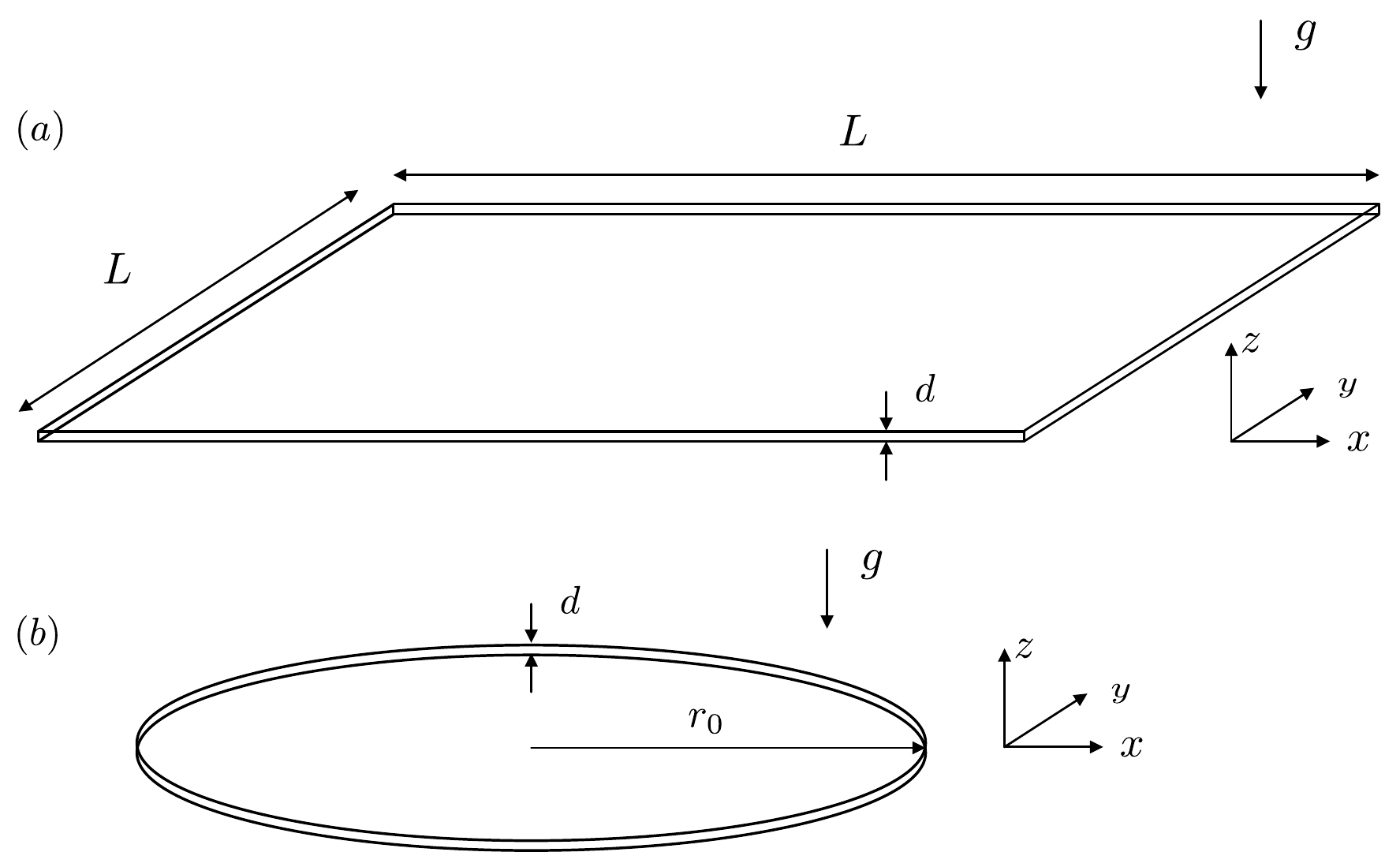}
\end{center}
\caption{Schematics of the two domains used for the numerical simulations of the Boussinesq equations. The Cartesian coordinates $(x,y,z)$ are in the directions shown and gravity acts in the direction opposing $z$.  (a)~The box domain with a square planform of side length $L$ and a depth $d$ with an aspect ratio of $\Gamma = L/d = 100$. The sidewall boundary conditions are periodic and the bottom and top walls are no-slip surfaces. This domain was used to generate a state of spiral defect chaos, with a sample flow field shown in Fig.~\ref{fig:bsqtemp}. (b) The cylindrical domain of radius $r_0$ and depth $d$ with an aspect ratio of $\Gamma = r_0/d = 40$. All material surfaces are no-slip boundaries and the sidewalls are heated as part of the procedure to develop a giant rotating spiral as described in the text. A sample flow field image is shown in Fig.~\ref{fig:gspiral} (left). Both schematics are drawn to scale and are shown slightly tilted with respect to the horizontal for perspective.}
\label{fig:schematics}
\end{figure}

The Prandtl number of the fluid $\text{Pr}= {\nu}/{\kappa}$ is the ratio of the momentum diffusivity to the thermal diffusivity. The connection between $\text{Pr}$ and the rescaled Prandtl number $\sigma$ used in the generalized Swift Hohenberg equation is described in Appendix A. The Prandtl number is inversely related to the magnitude of the mean flow~\cite{cross:1984,newell:1990:prl} which has been shown to have a significant effect upon the state of spiral defect chaos~\cite{assenheimer:1994,chiam:2003,karimi:2012,bodenschatz2000}. In the numerical simulations presented here, we use $\text{Pr}=1$ which is typical of the compressed gases often used in Rayleigh-B\'enard convection experiments~\cite{bodenschatz2000,morris1993spiral}.  The aspect ratio of the domain $\Gamma$ is the ratio of the lateral extent of the convection layer to its depth.  We have used two different geometries in our exploration reported here: a periodic box domain with $\Gamma=100$ to study spiral defect chaos and a cylindrical domain with $\Gamma=40$ to study a single rotating giant spiral.  Schematics of these two domains are shown in Fig.~\ref{fig:schematics}(a) and Fig.~\ref{fig:schematics}(b) for the box and cylindrical domains, respectively. We note that the spatial scale in this figure is different from the one used for Fig. \ref{fig:psi-theta} by a factor of $1/q_c$, as later detailed in the text and Appendix \ref{sec:ap}. %In Fig.~\ref{fig:schematics} the schematics are drawn to scale.

For the box domain, we used periodic boundary conditions at all the sidewalls while the bottom and top walls are no-slip surfaces.  For the thermal driving we used $\epsilon = 0.7$. In this case, our intention was to study the state of spiral defect chaos in a domain where the effects of the sidewall boundary conditions were reduced. For a box geometry with a square platform it is typical to define the aspect ratio as $\Gamma=L/d$, where $L$ is the length of the side of square domain. Using this convention we have $\Gamma=100$ for the results presented here for the box domain. We used initial conditions composed of small random thermal perturbations of magnitude $\delta T = 0.01$ to an otherwise quiescent layer of fluid. We then evolved the dynamics forward in time for approximately 930 time units to allow initial transients to decay.

\begin{figure}[htp]
	\centering
    \begin{subfigure}[b]{0.45\textwidth}
    \includegraphics[width=\textwidth]{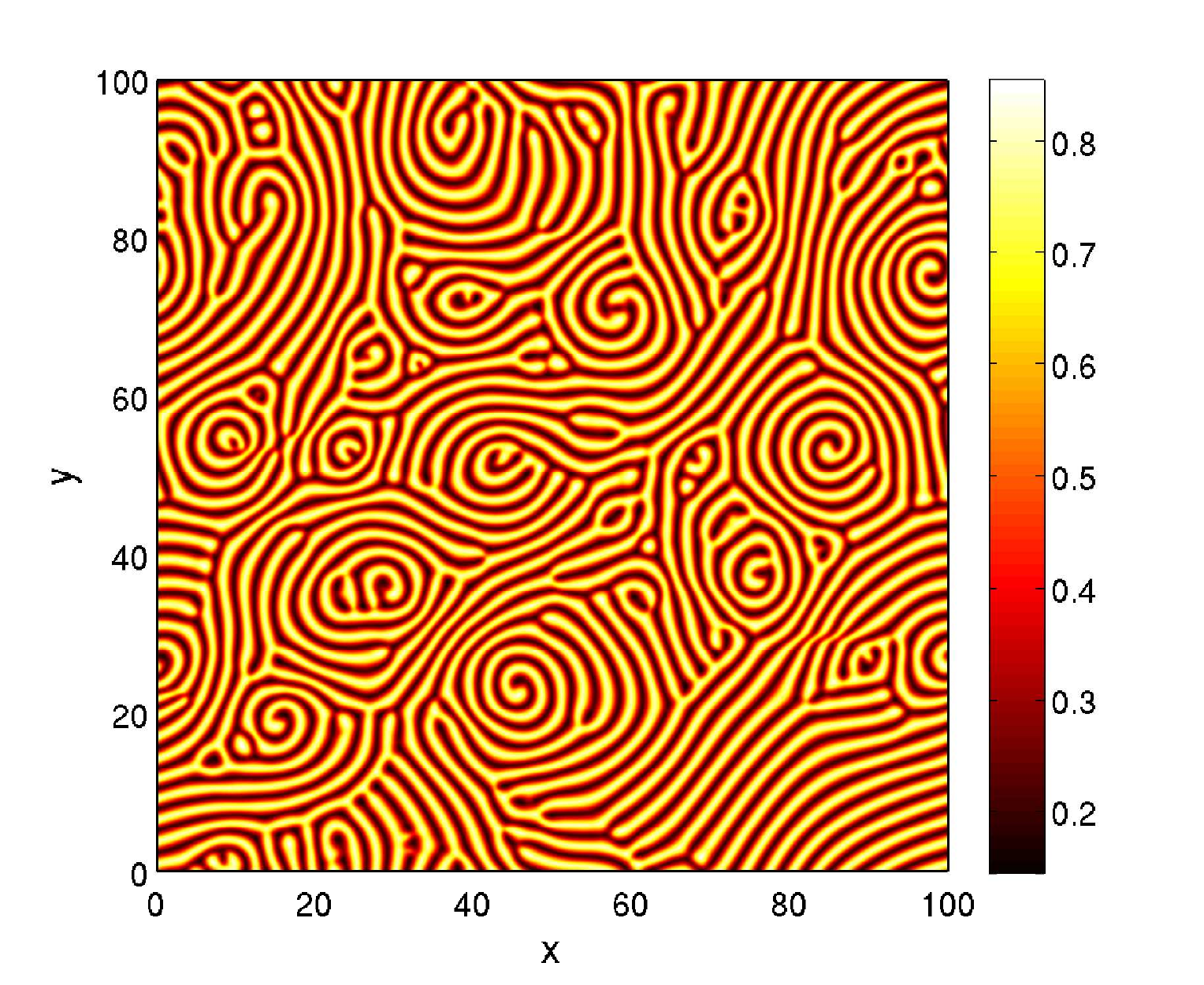}
    \end{subfigure}
    %\hspace{3mm}
    \begin{subfigure}[b]{0.45\textwidth}
    \includegraphics[width=\textwidth]{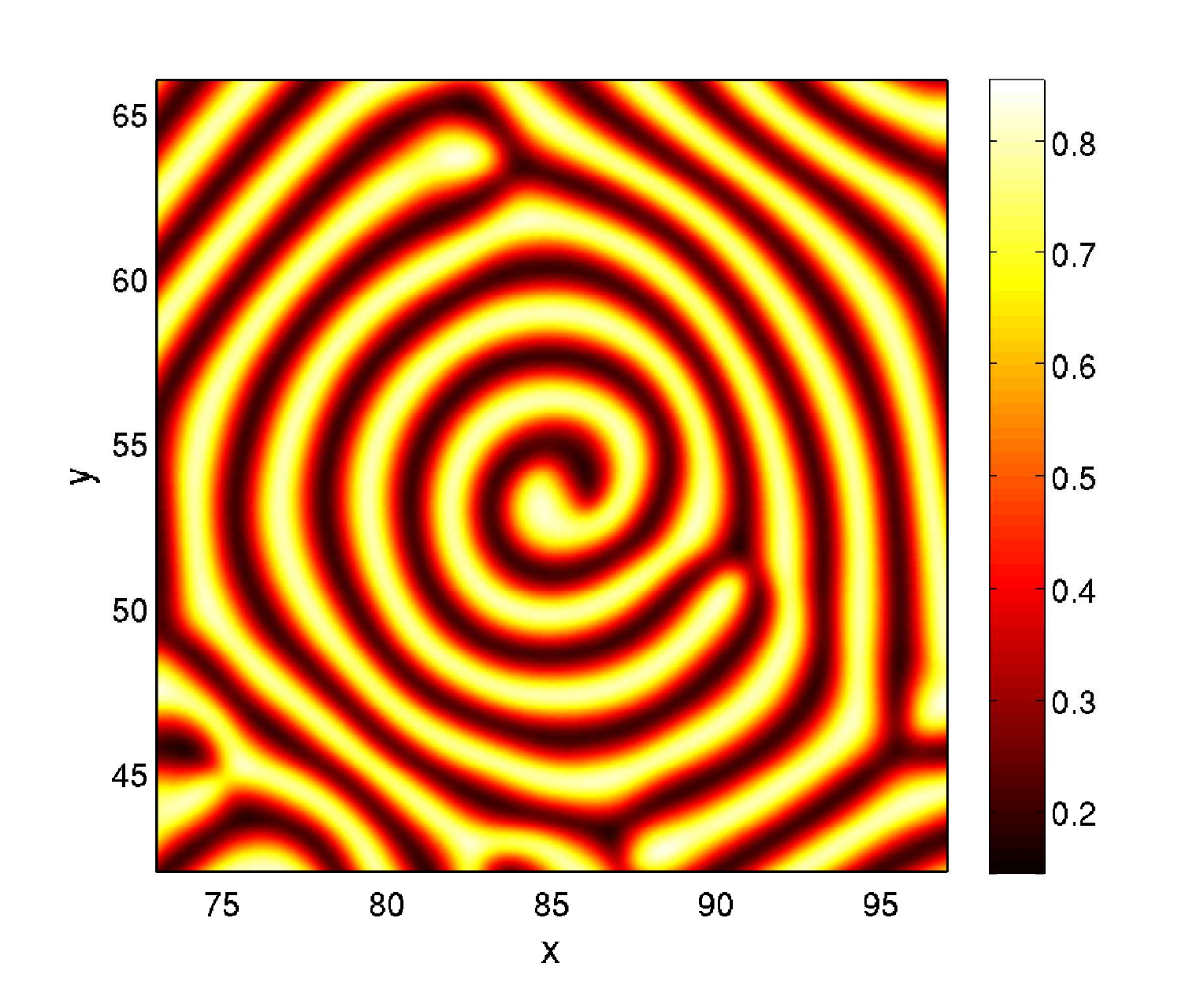}
    \end{subfigure}
    \caption{Left: Temperature field at the mid plane of the convection cell obtained by integrating the Boussinesq fluid model in time with periodic boundary conditions. The convection cell is a box domain with $\Gamma=100$, $\epsilon=0.7$ and $\text{Pr}=1$. The temperature field is shown at time $t=914.69$. Right: A close-up view of a rotating spiral.}
	\label{fig:bsqtemp}
\end{figure}

We note that this duration of time is less than a nondimensional horizontal heat diffusion time $\tau_h$ which is often used as a rough benchmark for determining the length of time required for a simulation to achieve a sufficient reduction of transients~\cite{cross:1984}. $\tau_h$ is the amount of time required for heat to diffuse from the center of the domain to a sidewall. For the box domain this yields $\tau_h \!=\! (L/2)^2 \!=\! 2500$. A simulation of this duration requires significant computational expense. We found that a duration of $930$ time units was sufficient to establish a steady state of spiral defect chaos. We are interested in the instantaneous features of the patterns, in particular in the features of the relatively short lived spiral structures, and not in the long time statistics of the global pattern dynamics. As a result, we anticipate that a time of $930$ time units is sufficient to study the mean flow field and the azimuthal flows that are generated around the spiral structures. An example flow field from a numerical simulation is shown in Fig.~\ref{fig:bsqtemp}.

In order to study a single rotating spiral we used a cylindrical domain of aspect ratio $\Gamma \! = r_0/d \! = \! 40$, where $r_0$ is the radius of the domain. To generate a large spiral in this domain we follow the approach used in the experiment of Plapp \emph{et al.}~\cite{plapp1998}. We initialize the simulation by starting with a quiescent layer of fluid where the lateral sidewalls are slightly heated while the thermal driving of the layer is just above threshold at $\epsilon = 0.054$. Specifically, the temperature at the sidewalls are set to the constant value of $T=0.1$ for all $z$, i.e., a hot sidewall boundary condition. The hot sidewall creates an up-flow at the wall which initializes the formation of a curved convection roll that aligns with the sidewall boundary. We then evolve the system forward in time for approximately $500$ time units; during this time curved convection rolls grow inward towards the geometric center of the domain, resulting in a stable and stationary target pattern. We next restart the simulation further from threshold with $\epsilon = 0.405$ and allow it to evolve for approximately $300$ time units. During this time, the center of the target pattern slowly drifts away from the geometric center of the domain to yield a stationary and time-independent skewed target-like pattern. We then restart the simulation further from threshold with $\epsilon = 0.464$ and let the system evolve for another $800$ time units. This causes the center of the target-like pattern to drift further from the geometric center of the domain where the pattern eventually undergoes a complex transition of instabilities that eventually yield the giant one-armed spiral with a single dislocation as shown in Fig.~\ref{fig:gspiral}. Both the giant spiral and the dislocation are rotating in the clockwise direction for these results.  This procedure appears to be a flexible and reliable way to generate giant spirals. However, the specific parameters and sequence we used were determined by trial and error with the goal of generating a giant spiral and are by no means meant to describe a unique procedure.

All of our numerical simulations of Eqs.~(\ref{eq:momentum})--(\ref{eq:mass}) were conducted using the high-order, highly parallelized, and open-source spectral element solver nek5000~\cite{nek5000,fischer1997,deville2002}.  The code uses a semi-implicit operator splitting approach that is third-order accurate in time and converges exponentially in space. A hallmark of the approach is its geometric flexibility while also permitting explorations of large spatially extended systems. The nek5000 solver has been used to explore a broad range of fluids problems~\cite{deville2002}. More details regarding its use to study spatiotemporal chaos in Rayleigh-B\'enard convection can be found in Refs.~\cite{paul2003pattern,scheel:2006,jayaraman:2006,paul:2007,karimi:2012}.

\section{Azimuthal flows in the chaotic regime}
\label{sec:azi}

We address in this section the extent to which the asymptotic results of Sec. \ref{sec:vel} can shed some light on the role of hydrodynamic flows on spiral defect chaos. There are a number of factors that preclude a precise comparison between these analytic predictions and our numerical results. First, the typical size of a spiral in the chaotic state is relatively small (a few rolls), making a determination of the asymptotic decay of the azimuthal velocity questionable. Second, the results of Sec. \ref{sec:vel} have been derived for the generalized Swift-Hohenberg model, and they exhibit a strong dependence on the damping parameter $c^{2}$. This makes a comparison with results from the Boussinesq model difficult as this parameter is largely phenomenological, although it has been estimated for the case of no-slip boundary conditions (see Ref. \cite{manneville1984modelisation} and Appendix \ref{sec:ap}). Third, we have not explored the dynamics using the Boussinesq model for the case of free-slip (stress-free) boundary conditions. In this case $c^{2} = 0$, and the logarithmic dependence of Eq. (\ref{eq:rlogr}) might be apparent, and with it strong mean flows and interactions between spirals. Nevertheless, we will argue that the azimuthal velocity field within a given spiral depends strongly on the cutoff radius $r_{b}$ for small values of $c^{2}$, thereby providing a mechanism for the hydrodynamic interaction of spirals.

We first use a chaotic configuration obtained from the solution of the the generalized Swift-Hohenberg model, to analyze the $r$-dependence of the azimuthal velocity $v_\varphi$ as given in Eqs.~(\ref{eq:bessel})-(\ref{eq:rlogr}). We use the same spiral configuration of Fig. \ref{fig:psi} obtained with the following values of model parameters: $\epsilon = 0.7, g_{m} = 50, c^{2} = 2$, and $\sigma = 1$. We then compute the corresponding $v_\varphi$ from Eq. (\ref{eq:vort}) from the instantaneous velocity field, for a range of values of $c^{2}$. That is, by setting the time derivative of Eq. (\ref{eq:vort}) to zero [or equivalently, from the curl of Eq. (\ref{eq:stokes})], we obtain the velocity field in the Stokes limit for various values of $c^2$ from the same order parameter $\psi$ configuration. This way, we are able to follow the evolution of $v_\varphi$ as a function of $c^2$ only, and evaluate if the transition from the $-r \text{ln} (r/r_b)$ behavior based on Eq. (\ref{eq:rlogr}) to the damped profiles of Eq. (\ref{eq:bessel}) is observed. Note, however, that for these parameter values we do not observe spirals in the simulation transients when $c^{2} \leq 0.1$; rather, we observe target defects (see also discussions in Sec. \ref{sec:adv}). 

\begin{figure}[htp]
	\centering
    \begin{subfigure}[b]{0.28\textwidth}
    \includegraphics[width=\textwidth]{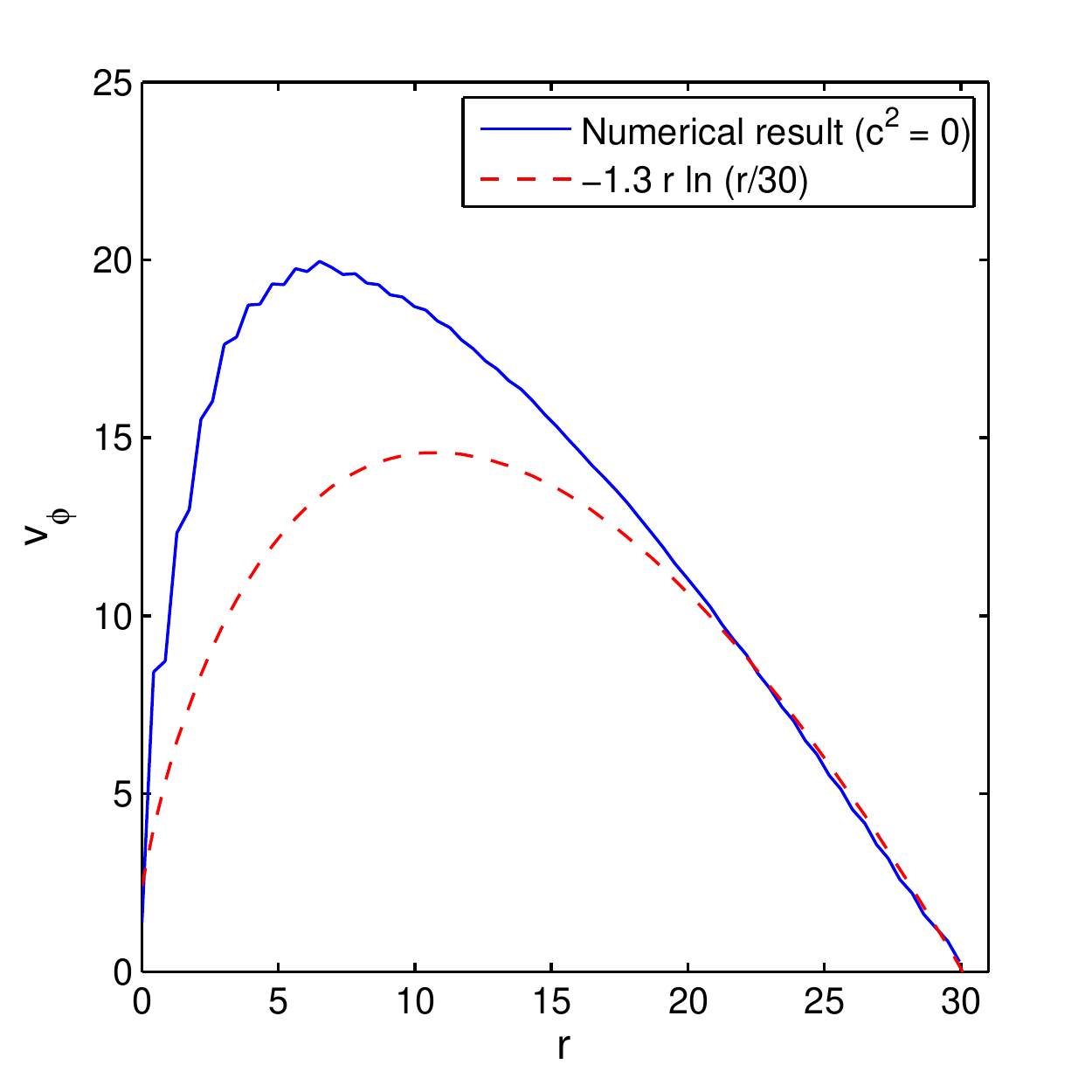}
    \end{subfigure}
    %\hspace{3mm}
    \begin{subfigure}[b]{0.28\textwidth}
    \includegraphics[width=\textwidth]{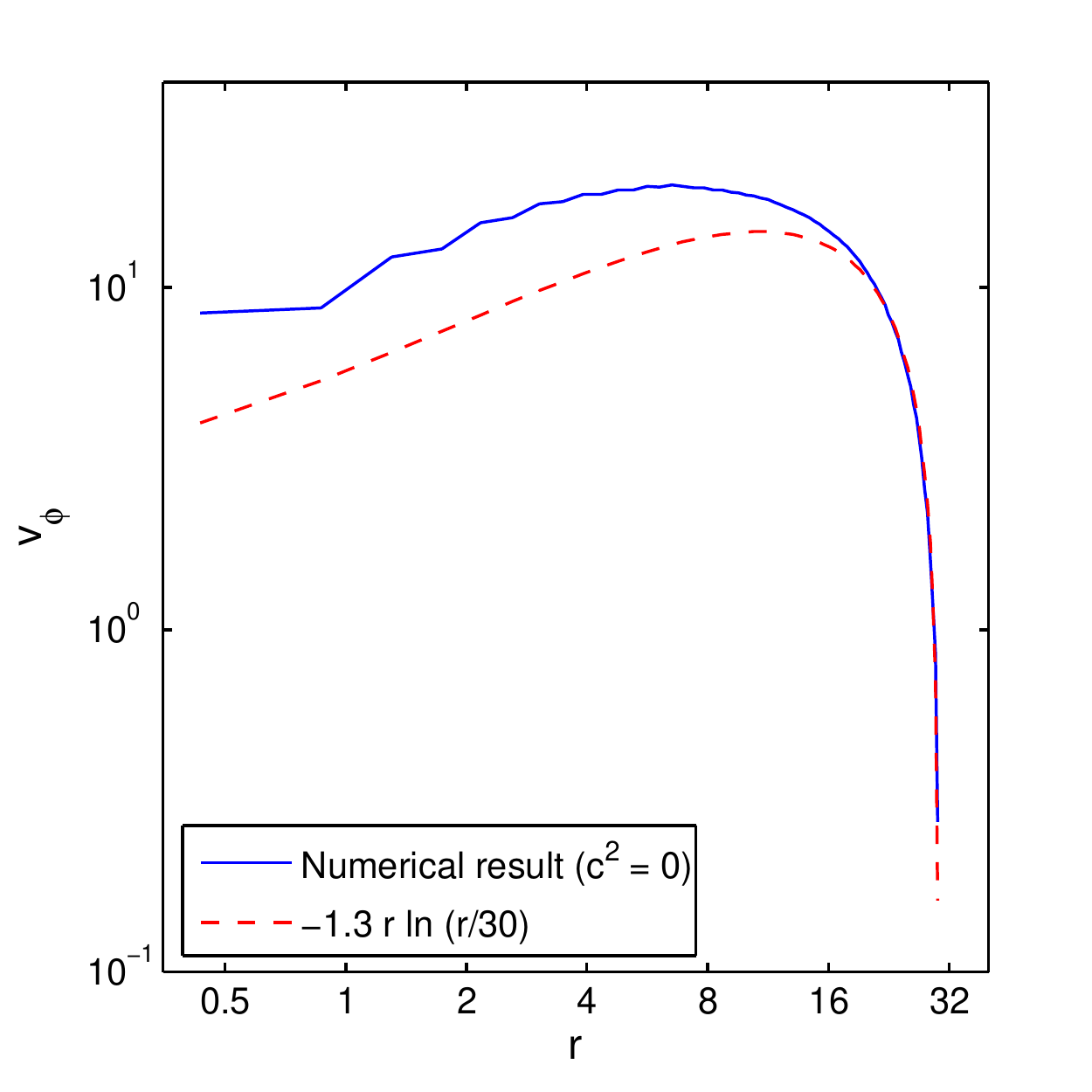}
    \end{subfigure} \\
    \begin{subfigure}[b]{0.28\textwidth}
    \includegraphics[width=\textwidth]{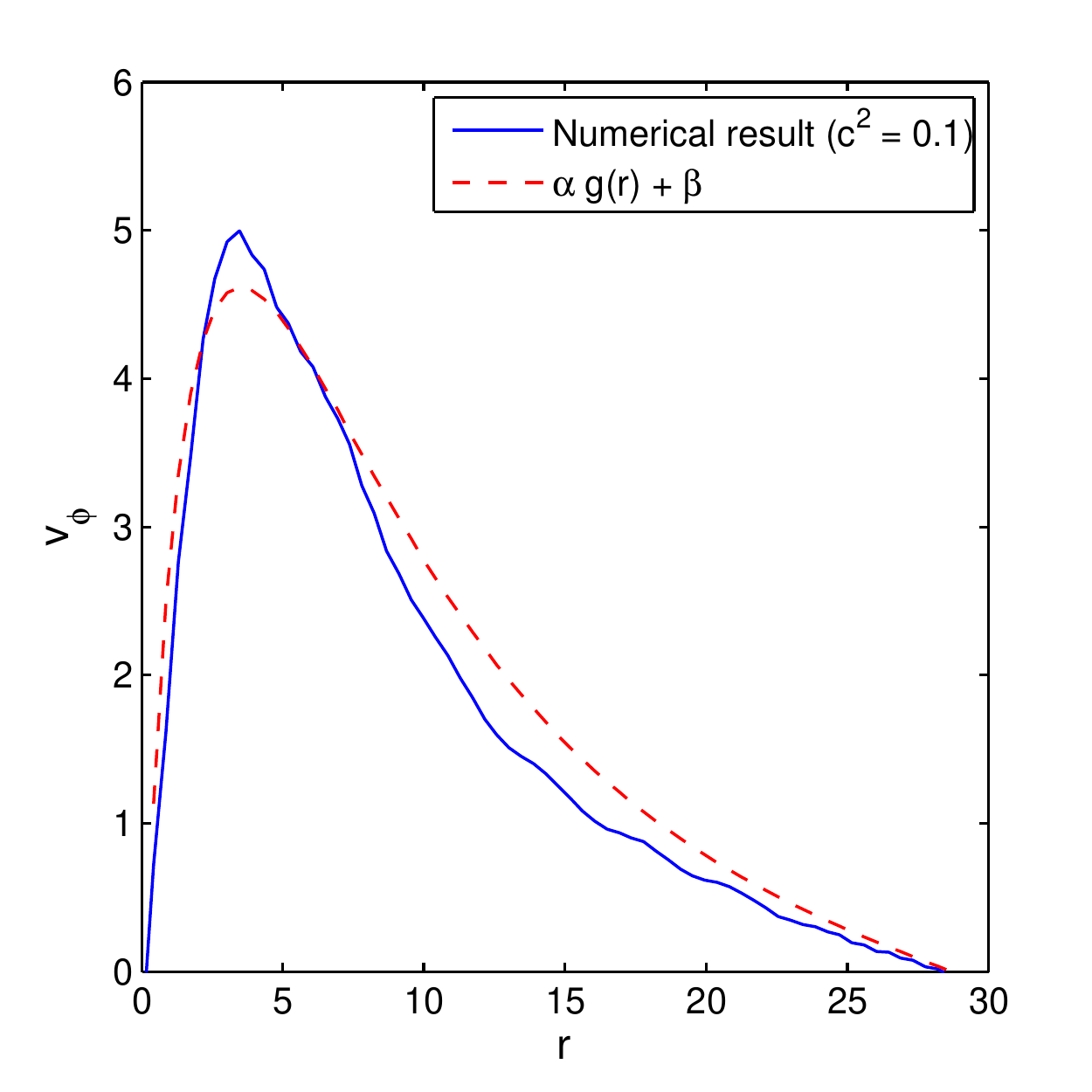}
    \end{subfigure}
    %\hspace{3mm}
    \begin{subfigure}[b]{0.28\textwidth}
    \includegraphics[width=\textwidth]{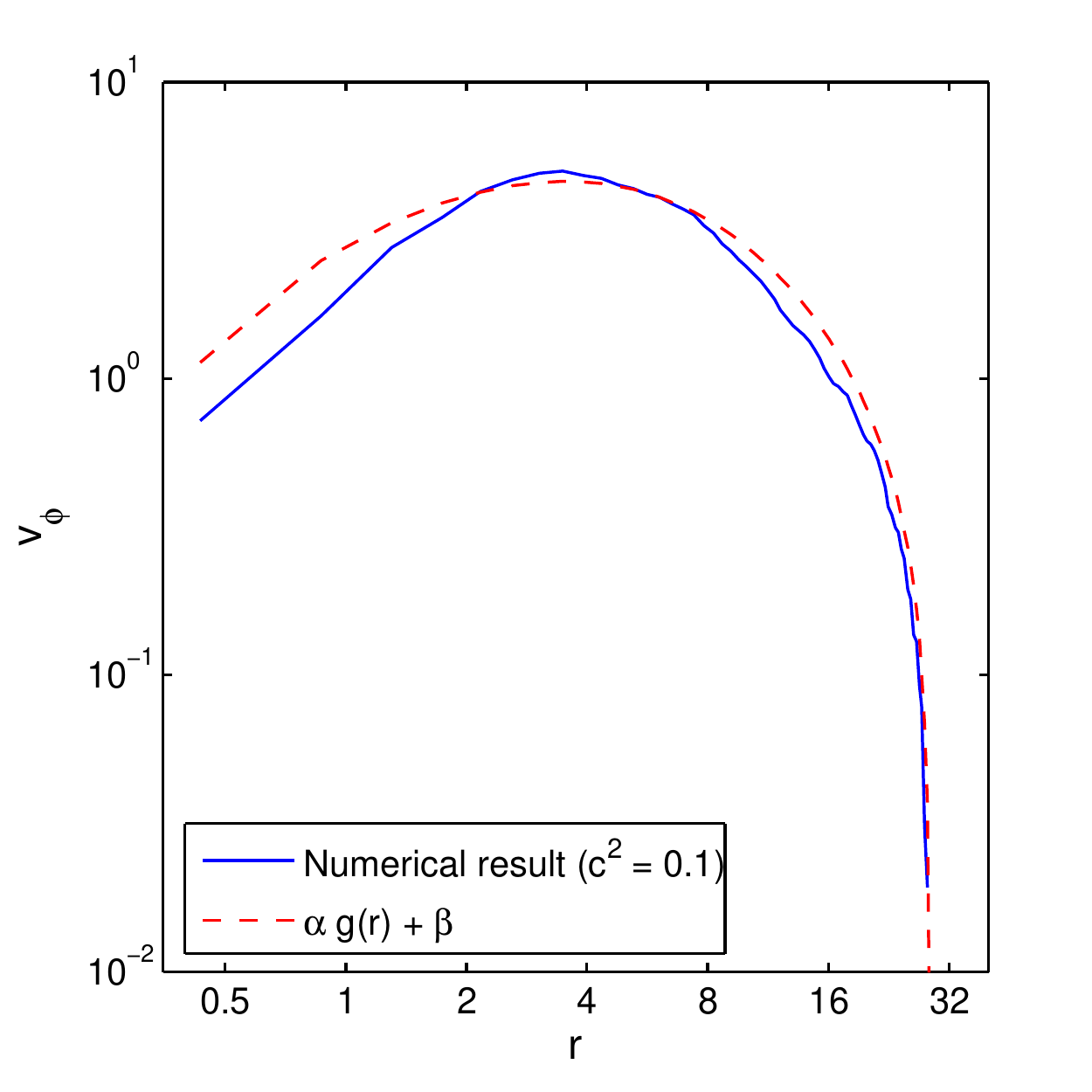}
    \end{subfigure} \\
    \begin{subfigure}[b]{0.28\textwidth}
    \includegraphics[width=\textwidth]{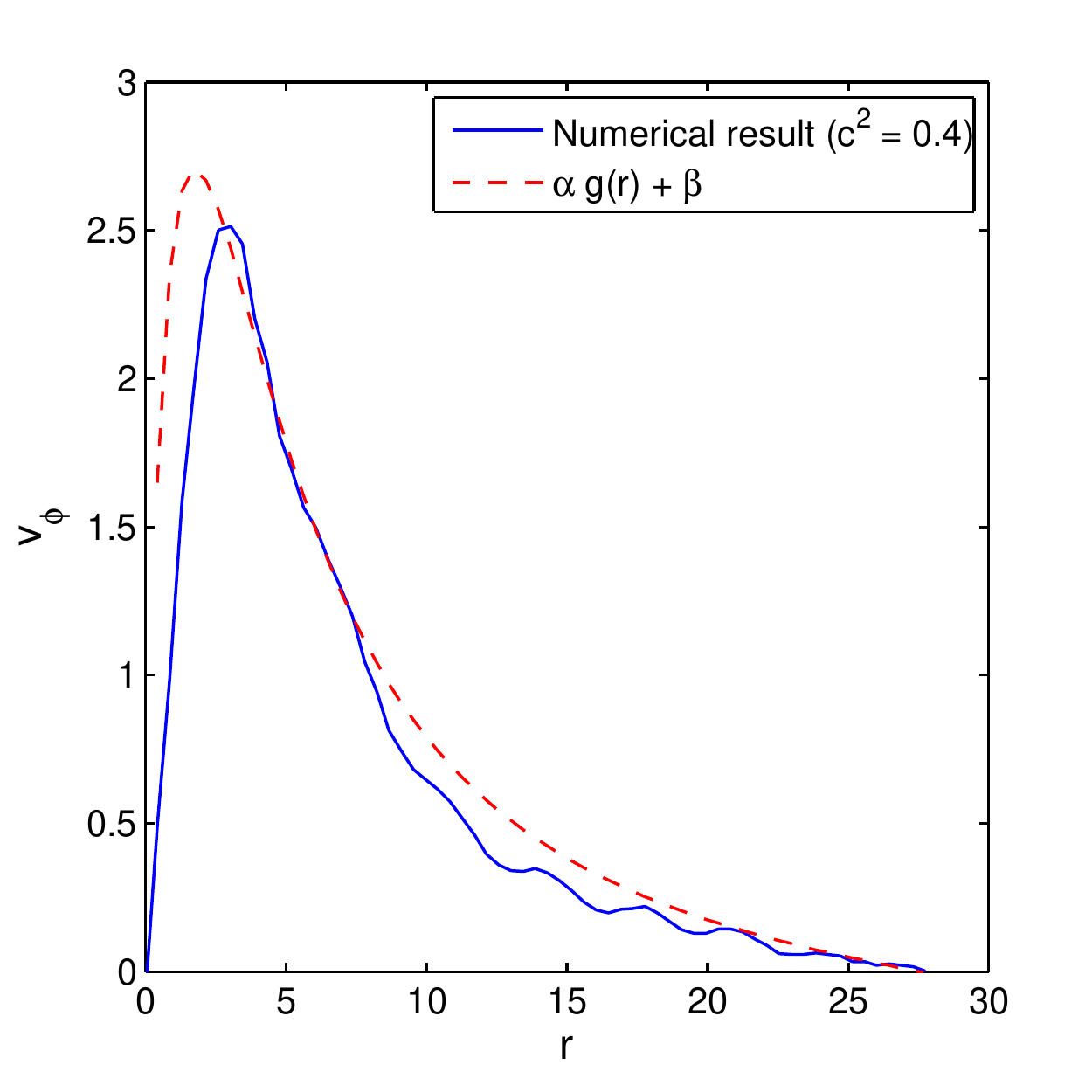}
    \end{subfigure}
    %\hspace{3mm}
    \begin{subfigure}[b]{0.28\textwidth}
    \includegraphics[width=\textwidth]{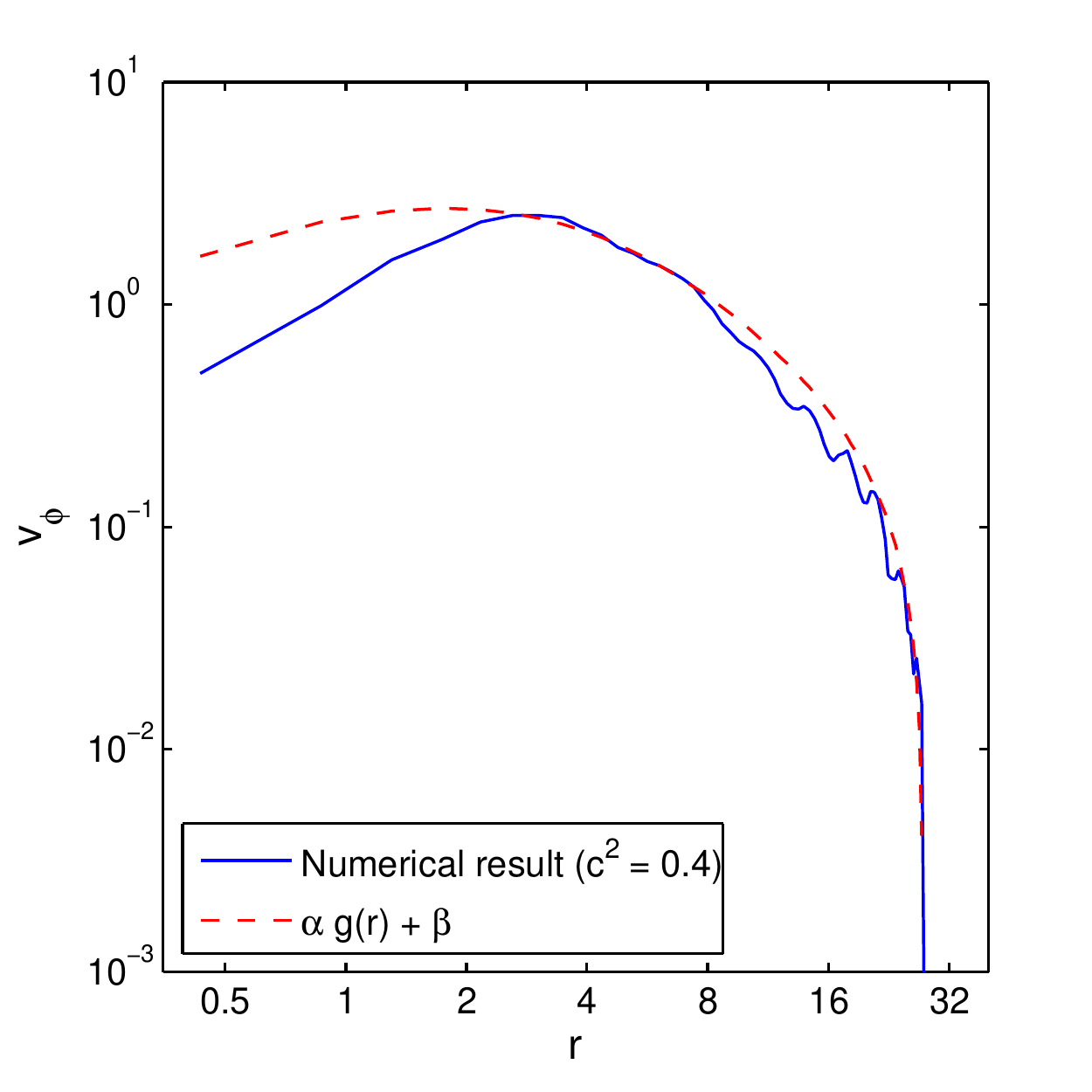}
    \end{subfigure} \\
    \begin{subfigure}[b]{0.28\textwidth}
    \includegraphics[width=\textwidth]{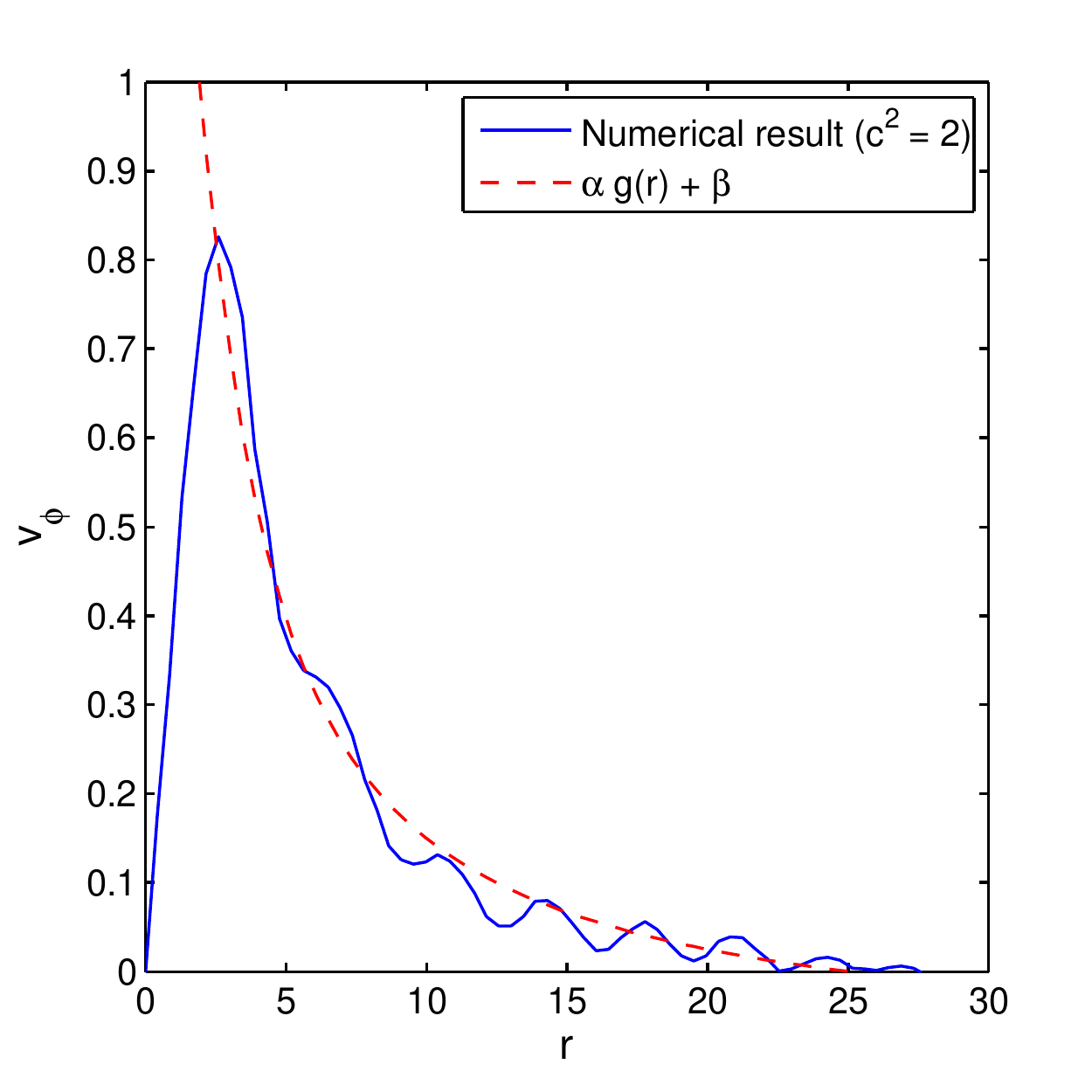}
    \end{subfigure}
    %\hspace{3mm}
    \begin{subfigure}[b]{0.28\textwidth}
    \includegraphics[width=\textwidth]{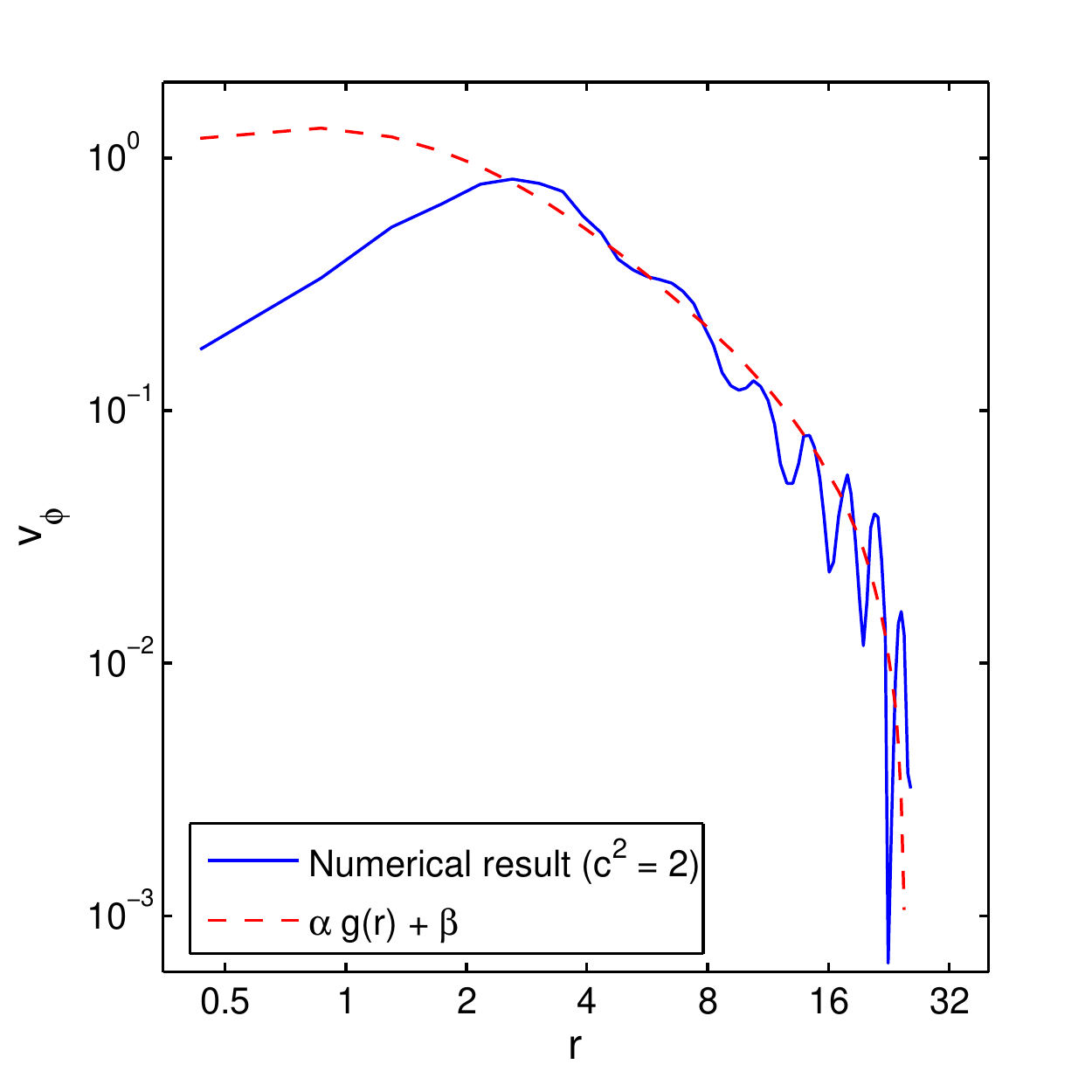}
    \end{subfigure}
    \caption{Azimuthal velocity for the spiral located at $(104,56)$ in Fig. \ref{fig:same} with $r=0$ at its core, for $g_m = 50$, $\sigma = 1$, and $\epsilon = 0.7$. Left column compares numerical results with our analytic predictions, and right column is in logarithmic scale. First row: $c^2 = 0$, using $v_\varphi = -1.3r\ln(r/30)$ for the analytic curve. Second row: $c^2 = 0.1$, using $\alpha = 5/\sigma$ and $\beta = -1.75$. Third row: $c^2=0.4$, using $\alpha = 5/\sigma$ and $\beta = -0.5$. Fourth row: $c^2=2$, using $\alpha = 5/\sigma$ and $\beta = -0.1$. For all the cases of $c^2 > 0$, $r_b = 35$ is used.}
	\label{fig:vphi}
\end{figure}

Figure \ref{fig:vphi} shows the azimuthal velocity $v_\varphi$ away from $r=0$ at the core of the spiral located at coordinate $(104,56)$ in Fig. \ref{fig:same}, up to the midpoint between this spiral and the other one at $(156,81)$. The figure shows the numerical solution for $c^2 = 0$, $c^2 = 0.1$, $c^2 = 0.4$ and $c^2 = 2$ (solid lines).  In order to compare the numerical solution with the analytic result of Eq. (\ref{eq:bessel}) with $c^{2} > 0$, we define
\begin{equation}
  g(r) = \frac{1}{c^2}\Bigg[ \frac{1}{r}+\Bigg(c K_1(r_b c) - \frac{1}{r_b}\Bigg)\frac{I_1(c\,r)}{I_1(r_b c)} - c K_1(c\,r) \Bigg],
\end{equation}
and fit the function $v_\varphi  = \alpha\,g(r) + \beta$, where $\alpha$ and $\beta$ are two fitting coefficients. The function $g(r)$ depends on two parameters, the damping parameter $c^2$ and a cutoff radius $r_b$ which is taken to be of the order of the spiral size.

For $c^{2} = 0$ and the two rotating spirals of Fig. \ref{fig:same}, we set $r_{b} = 30$, the mid point between the two spirals (we see that $v_\varphi$ changes sign approximately at $r=30$). The asymptotic relation $v_\varphi = -1.3 r\ln(r/30)$ is shown in the top two panels of Fig. \ref{fig:vphi} (dashed lines), where the constant $-1.3$ is the single fitting parameter. There is good agreement away from the core. Figure \ref{fig:vphi} also shows our results for $c^{2} > 0$. We have set $r_{b} = 35$ in all these cases, and fit the parameters $\alpha$ and $\beta$. From Eq. (\ref{eq:bessel}) we note that $\alpha = m\omega\epsilon g_m/6q_0^2\sigma$, where the only unknown is the angular frequency of rotation $\omega$. Using our current parameter values, we have $\alpha =  5.83\,\omega$, where $\omega$ is not known \emph{a priori}. The order of magnitude of the angular frequency will be further discussed in Sec. \ref{sec:adv}, where we find it to be on the order of $10^{-1}-1$. We have assumed that the constant $\alpha$ is largely independent of $c^{2}$, and chosen $\alpha = 5/\sigma$ for all the values of $c^{2} > 0$ in Fig. \ref{fig:vphi}, where the rescaled Prandtl number is $\sigma = 1$. Therefore the only fitting parameter used in Fig. \ref{fig:vphi} for $c^{2} > 0$ is the constant $\beta$.  We note that Eq. (\ref{eq:bessel}) is valid away from the core, and is obtained with the boundary condition of vanishing velocity at the core. Hence the constant $\beta$ can be rationalized as being related to the velocity near the core that should be used as a known boundary condition for the outer solution. The fitted $\beta$ value also contributes to the large radius at which $v_{\varphi}$ vanishes. For finite damping, the azimuthal velocity does not completely decay to zero, so that we need a negative $\beta$ to capture such effect. For small values of $c^{2}$, the value of $r_{b}$ is relatively easy to determine, and is closely related to $\beta$ in order to obtain a good fit. As $c^{2}$ increases, the velocity field decays quickly, and the fitting becomes less dependent on the value of cutoff $r_{b}$ as long as $r_b$ remains greater than the size of the spiral.

By increasing the damping coefficient from $c^2=0$ to $c^2 = 0.1$, the magnitude of the azimuthal velocity diminishes by a factor of four, and its asymptotic decay changes from convex to concave, as expected from Eq. (\ref{eq:bessel}). Figure \ref{fig:vphi} shows that our analytic prediction from Eq. (\ref{eq:bessel}) appropriately describes the behavior of $v_\varphi$ far enough from the spiral core for all the $c^2>0$ cases. We also note that in every case $v_{1\varphi}$ is sufficient to describe the form of the numerical curves, while $v_{2\varphi}$ and $v_{3\varphi}$ do not provide any major contribution to the observed results. 

As $c^2$ increases, the flow field becomes increasingly localized within the vicinity of the spiral cores, thereby reducing any interaction between velocity fields generated by different spirals. This is illustrated in Fig. \ref{fig:vphi} when comparing the cases of $c^2 = 0.4$ and $c^2 = 2$. 
%For both cases, Eq. (\ref{eq:bessel}) still describes the behavior of the angular velocity, using the same $\alpha = 5/\sigma$ parameter. For $c^2 = 0.4$, we use $\beta = -0.45$, and for $c^2 = 2$ we use $\beta = -0.1$. 
As the flow damping increases, the velocity becomes more short-ranged, and the magnitude of the azimuthal velocity decreases significantly: for $c^2 = 2$, the flow magnitude is only $4\%$ of the flow obtained for $c^2 = 0$. Note that while $r_b$ is of the order of the spiral's size, the velocity for larger values of $c^2$ reaches zero well before $r = 30$, and that due to the small size of the spirals we do not observe a $1/r$ decay at long distances. In this range of $c^{2}$,  the velocity field within each spiral is largely independent of the existence of other spirals, and does not depend on the value of $r_{b}$ \cite{cross1996theoretical}.

In summary, we have observed the transition of the azimuthal velocity  from a $-r\textrm{ln}(r/r_b)$ profile to the damped convex profiles when the damping coefficient $c^2$ increases, as suggested by our predictions in Sec. \ref{sec:vel}. As $c^2$ approaches zero, given the longer range of the flows the cutoff parameter $r_b$ has the same value as the spiral's radius when there are spirals of the same topological charge in the vicinity. By increasing $c^2$, as long as $r_b$ is greater than this radius, varying the cutoff makes little quantitative difference to the fits, since the azimuthal velocity decays quickly to zero. In addition, the azimuthal velocity field within a spiral strongly depends on the existence of neighboring spirals, and their presence affects the fit parameters $r_b$ and $\beta$. The topological charge of the spirals also plays a role in this observation, as will be discussed in Sec. \ref{sec:int} for the case of neighboring counter-rotating spirals.

\section{Discussion}

\subsection{Azimuthal flow between two counter rotating spirals}
\label{sec:int}

\begin{figure}[htp]
    \begin{subfigure}{0.32\textwidth}
        \centering
        \includegraphics[width=\linewidth]{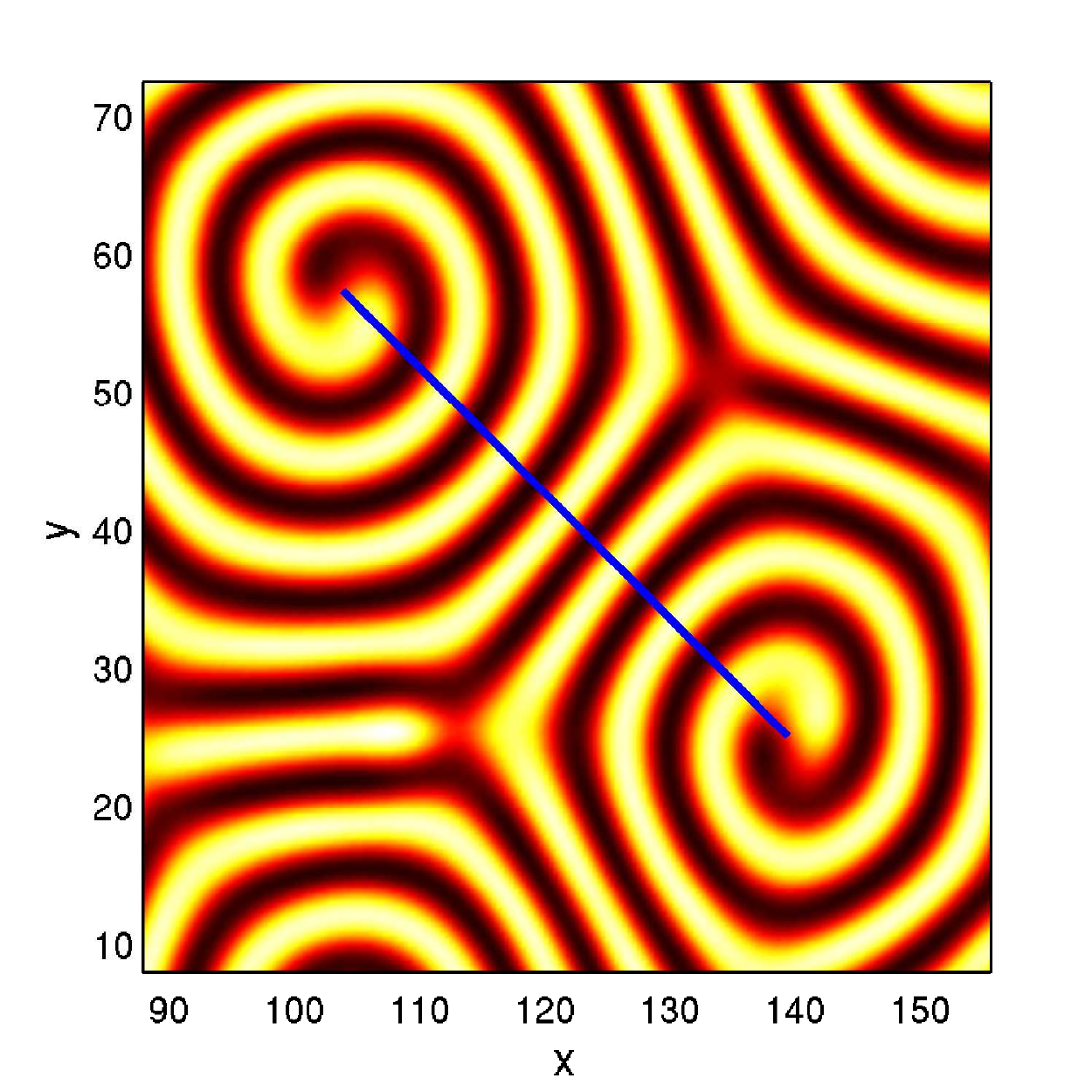}
    \end{subfigure}
    \begin{subfigure}{0.32\textwidth}
        \centering
        \includegraphics[width=\linewidth]{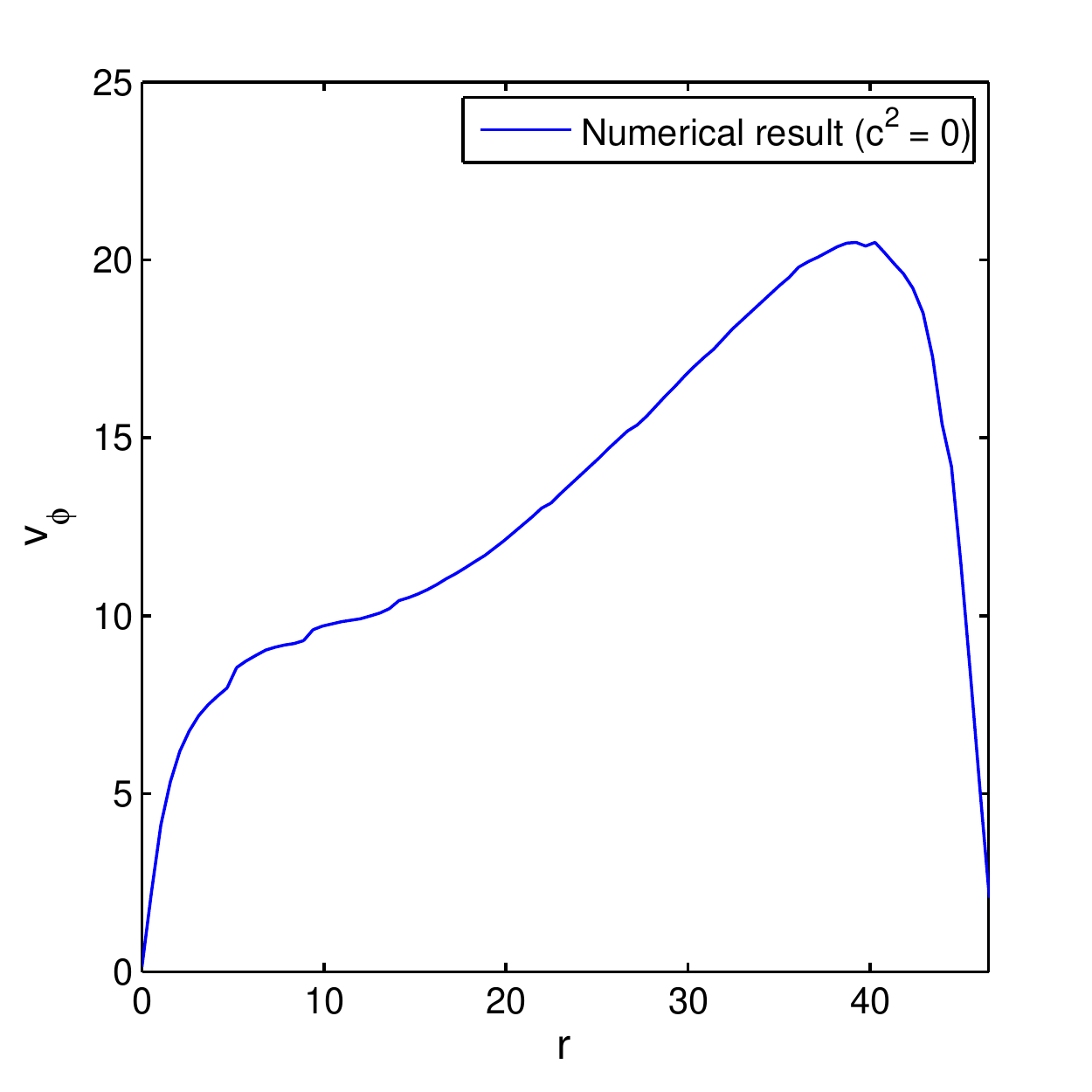}
    \end{subfigure}
    \begin{subfigure}{0.32\textwidth}
        \centering
        \includegraphics[width=\linewidth]{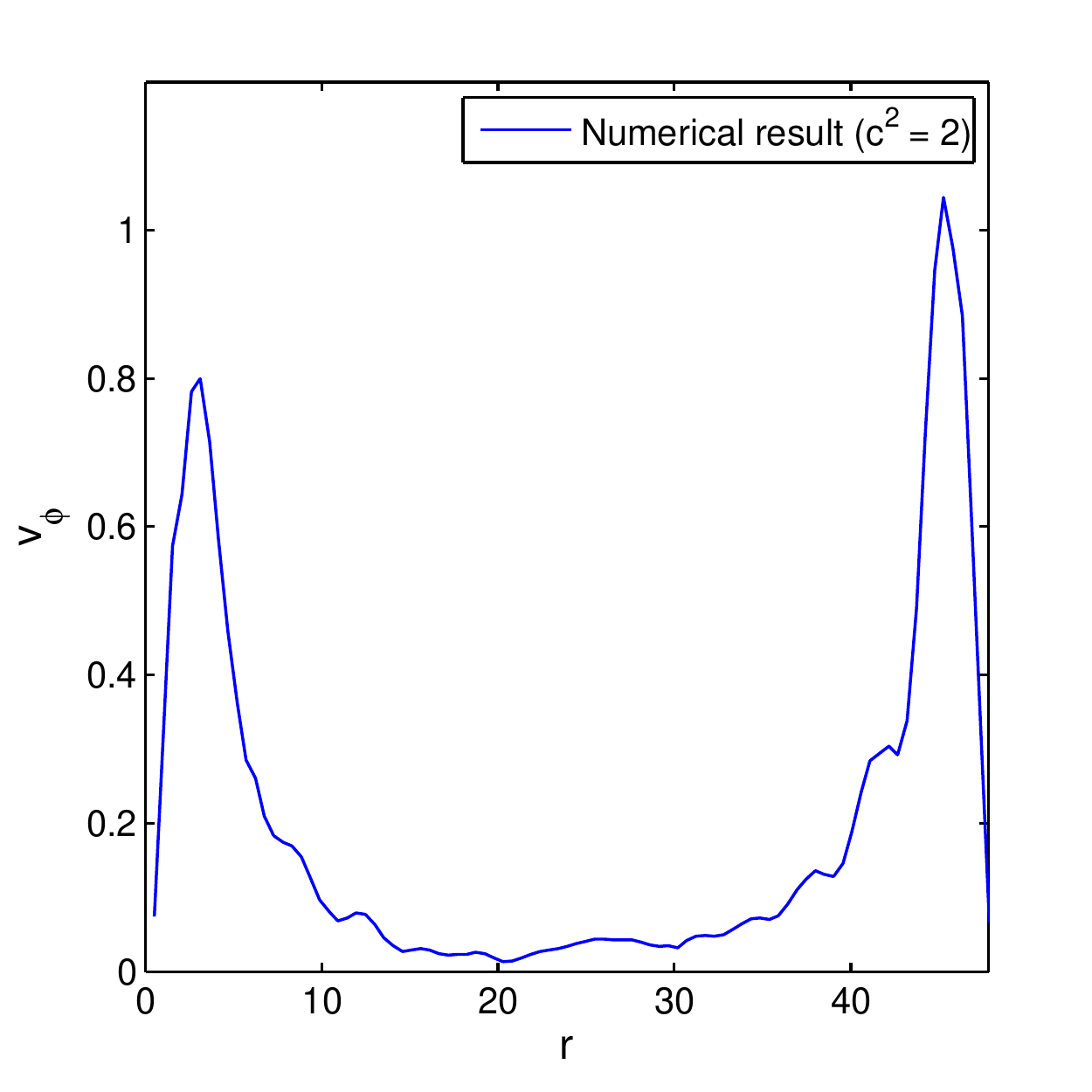}
    \end{subfigure}
    \caption{Azimuthal velocity between two spirals of opposite topological charge using the generalized Swift-Hohenberg equation. At $r = 0$ we find the core of the spiral located at $(104,56)$ from Fig. \ref{fig:same}, and at $r = 47$ the core of the spiral is located at $(139,25)$. Using this same order parameter configuration, we compute the instantaneous velocity for $c^2 = 0$ (middle panel) and $c^2 = 2$ (right panel), and plot the azimuthal velocity between the two spirals.}
    \label{fig:opposite}
\end{figure}

In the spiral chaos regime the flow field within each spiral depends on the spiral size, which in turns is determined by the presence of neighboring spirals and other defects through the cutoff parameter $r_b$. In particular, the decay of $v_\varphi$ with distance $r$ is faster than the asymptotic $1/r$. For stress free boundary conditions, $c^2 = 0$, or small damping (e.g., $c^2 = 0.1$) there is a strong and long ranged azimuthal velocity component spanning the entire spiral, which decays to zero at a scale determined by neighboring spirals of the same topological charge. For low damping, $r_b$ is approximately the spiral's radius. We present here an analysis of the flow between two counter rotating spirals (with opposite topological charge), as the flow would interact constructively along a line connecting them. Figure \ref{fig:opposite} shows the azimuthal velocity between two neighboring counter rotating spirals, centered at coordinates $(104,56)$ and $(139,25)$ in Fig. \ref{fig:psi}. We again use the same $\psi$ configuration shown in Fig. \ref{fig:psi} to compute the azimuthal velocities in the absence of inertia for two different values of the damping coefficient $c^{2}$. In the absence of damping, $c^{2} = 0$, the azimuthal velocity is nonzero in the region between the two spirals, as the vorticity generated by the two cores adds up constructively. For $c^2 = 2$, the flow once again becomes concentrated at each spiral, with small or no flow interaction between them.

\subsection{Advection versus roll unwinding in spiral dynamics}
\label{sec:adv}

We address here the possible relevance of the mean flows discussed to the chaotic state itself, based on the generalized Swift-Hohenberg model. It has been established that spiral defect chaos is only observed for a specific range of $c^2$ and scaled Prandtl number $\sigma$. For the parameter set used here, $g_m = 50$, $\epsilon = 0.7$ and $\sigma = 1\sim2$, spiral defect chaos has been found in the range $0.1 < c^2 < 5$ \cite{karimi2011exploring,xi1993spiral,cross1996theoretical}. For $c^2 > 4$, the leading-order Lyapunov exponent of the flow approaches zero \cite{karimi2011exploring}. In the opposite range of small damping, $c^2 \leq 0.1$, spiral defects are no longer observed while the system dynamics is chaotic. In our calculations, if $c^2 = 0$ the magnitude and range of the mean flow increase significantly and we are only able to achieve spiral defect chaos for this free-slip condition by reducing $g_m$ significantly to $g_m \sim 5$ (or, similarly, by increasing $\sigma$).

We examine here the relative contribution to the overall time variation rate $\partial_{t} \psi$ from the mean flow advection $\mathbf{v}\cdot\bm{\nabla}\psi$, and the diffusive pattern dynamics given by the RHS of Eq.~(\ref{eq:sh}), leading to roll unwinding \cite{cross1984convection,cross1996theoretical}. The magnitude of the latter depends on the value of the local wavenumber when it is maintained away from the critical value (i.e., wavevector frustration) \cite{cross1996theoretical,re:huang07}, and also from the curvature of the rolls \cite{vitral2019role}. Both contributions have been estimated theoretically \cite{cross1986wavebumber,cross1996theoretical}. 

\begin{figure}[ht]
	\centering
    \begin{subfigure}[b]{0.45\textwidth}
    \includegraphics[width=\textwidth]{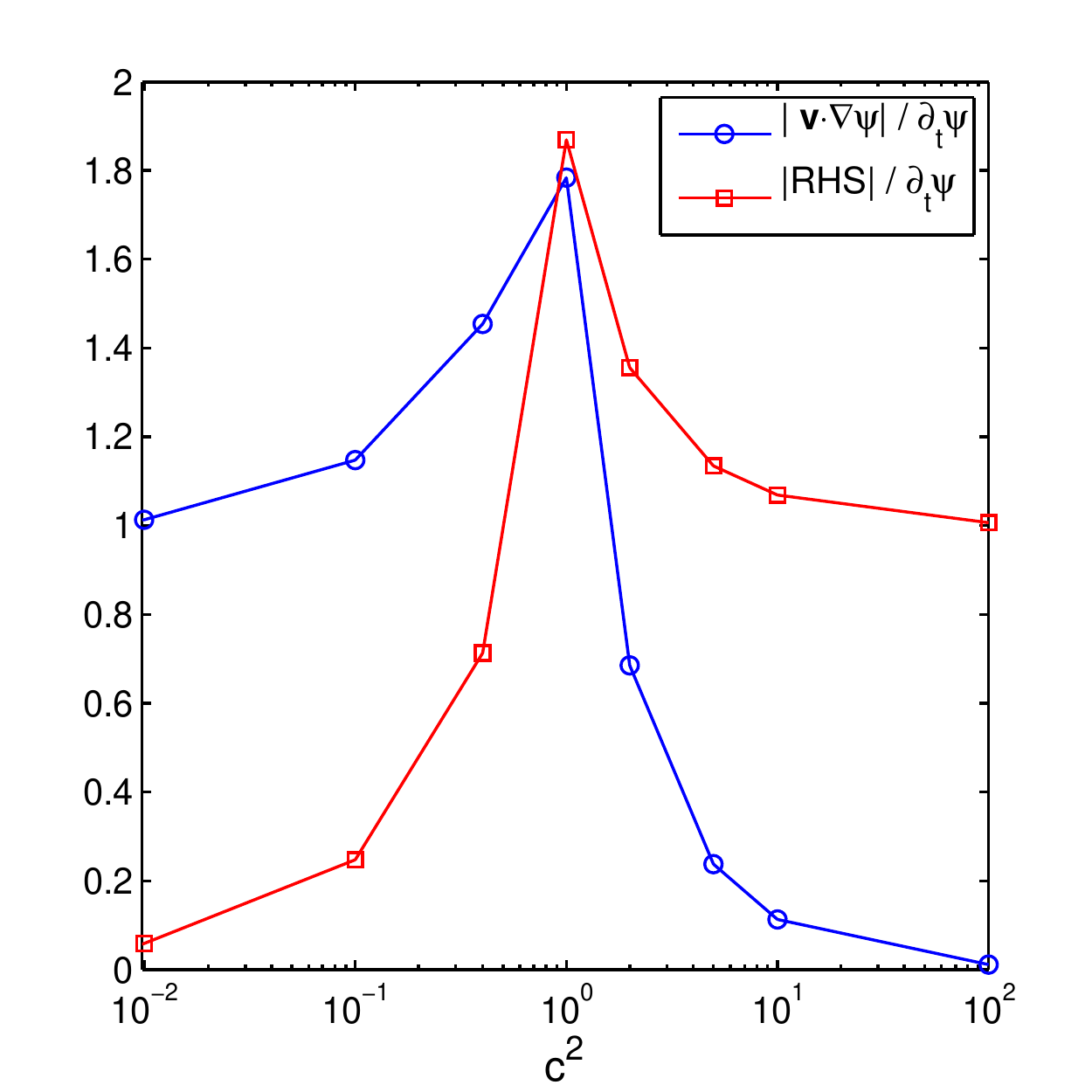}
    \end{subfigure}
    %\hspace{3mm}
    \begin{subfigure}[b]{0.45\textwidth}
    \includegraphics[width=\textwidth]{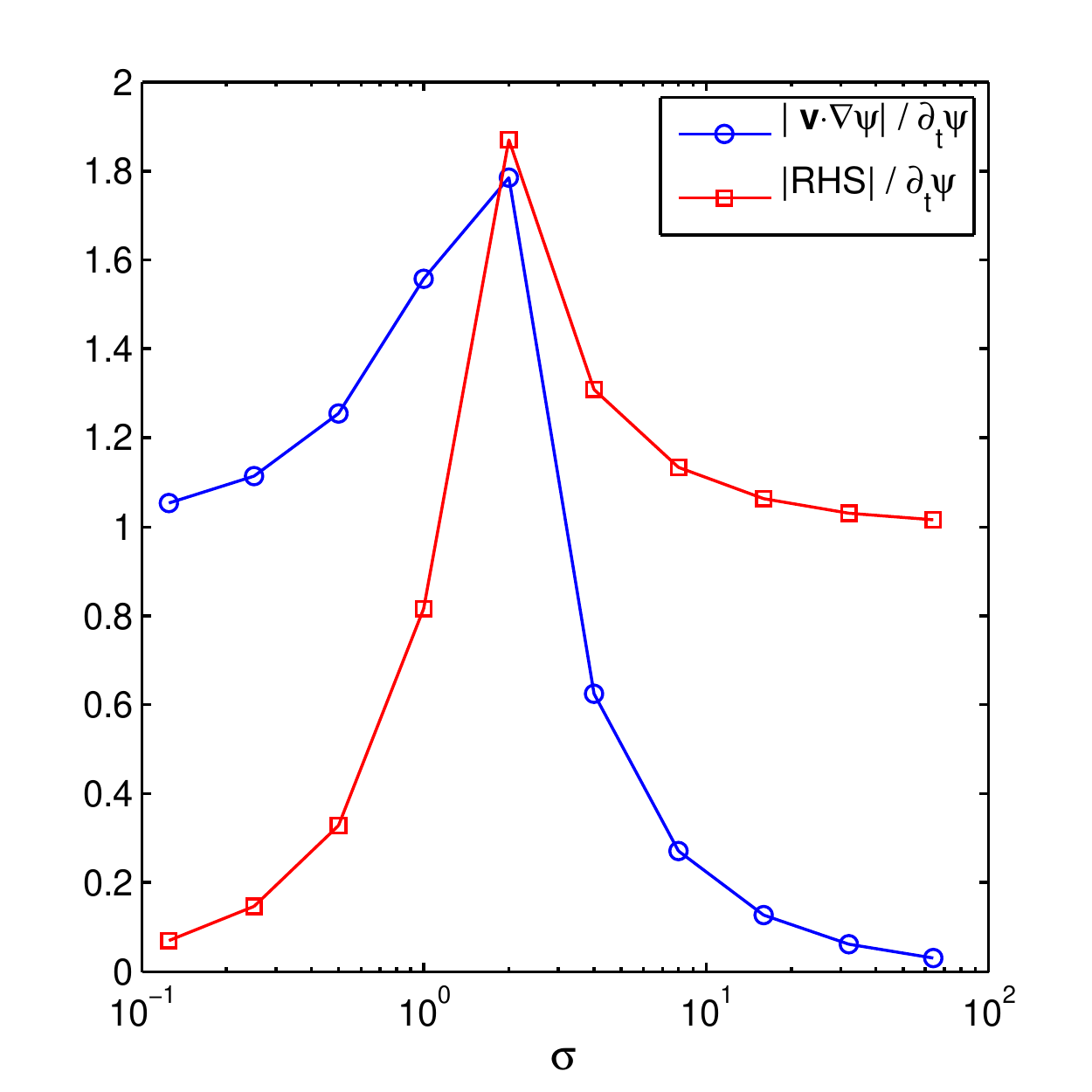}
    \end{subfigure}
    \caption{Ratios of advection and roll unwinding [the RHS of Eq. (\ref{eq:sh})] to the overall time variation $\partial_t\psi$, as a function of the damping coefficient $c^2$ and the rescaled Prandtl number $\sigma$. Values are computed based on the spiral located at $(104,56)$ in Fig. \ref{fig:psi}, with $\epsilon = 0.7$ and $g_m = 50$. The blue line with circle symbols shows the ratio for the spiral's characteristic advection contribution ($\mathbf{v}\cdot\bm{\nabla}\psi$), and the red line with square symbols shows the ratio for its characteristic roll unwinding contribution (RHS). Left: Ratios are plotted as a function of $c^2$, for $\sigma = 2$. Right: Ratios are plotted as a function of $\sigma$, for $c^{2} = 1$.
    }
	\label{fig:hydro}
\end{figure}

We use the same configuration of the order parameter field shown in Fig. \ref{fig:same}, and analyze the flow field around the spiral with core located at $(104,56)$. We obtain the velocity field by solving Eq. (\ref{eq:vort}) with the time derivative set to zero, and for a range of values of $c^{2}$ and $\sigma$. The overall time variation $\partial_{t} \psi$, advection $\mathbf{v}\cdot\bm{\nabla}\psi$, and the relaxational part (i.e., the RHS of Eq. (\ref{eq:sh}) yielding diffusive dynamics) oscillate nonuniformly as a function of the radial coordinate $r$. We extract the characteristic magnitude of each quantity by finding its maximum absolute value between $r = 5$ (away from the core) and $r = 28$ (the approximate radius of the spiral). Other measures, such as choosing the values from the first peak of these functions, lead to similar results. Our results are shown in Fig. \ref{fig:hydro} for a range of values of $c^{2}$ for fixed $\sigma = 2$, and also as a function of $\sigma$ for fixed $c^{2} = 1$. As described in Appendix \ref{sec:ap}, the value of $\sigma = 2$ corresponds to a Prandtl number of $\text{Pr} = 1$ (consistent with the CO$_2$ experiments of Ref. \cite{morris1993spiral}). We have conducted calculations across the range $0 \leq c^2 \leq 100$, and find that advection and diffusion contributions are of similar value around $c^2 = 1$. When rescaling the critical wavenumber to $q_0 = 1$, the value $c^2 = 1$ is the one estimated for no-slip boundary conditions on the cell's plates, as detailed in Ref. \cite{manneville1984modelisation} and Appendix \ref{sec:ap}. Next, we fix $c^2 = 1$ and compute the same ratios for a range of $0.125 \leq \sigma \leq 64$. Interestingly, both advection and diffusion contributions have approximately the same magnitude at $\sigma = 2$ (i.e., at the experimentally used Prandtl number $\text{Pr} = 1$).

\begin{figure}[ht]
    \begin{subfigure}{0.45\textwidth}
        \centering
        \includegraphics[width=0.9\linewidth]{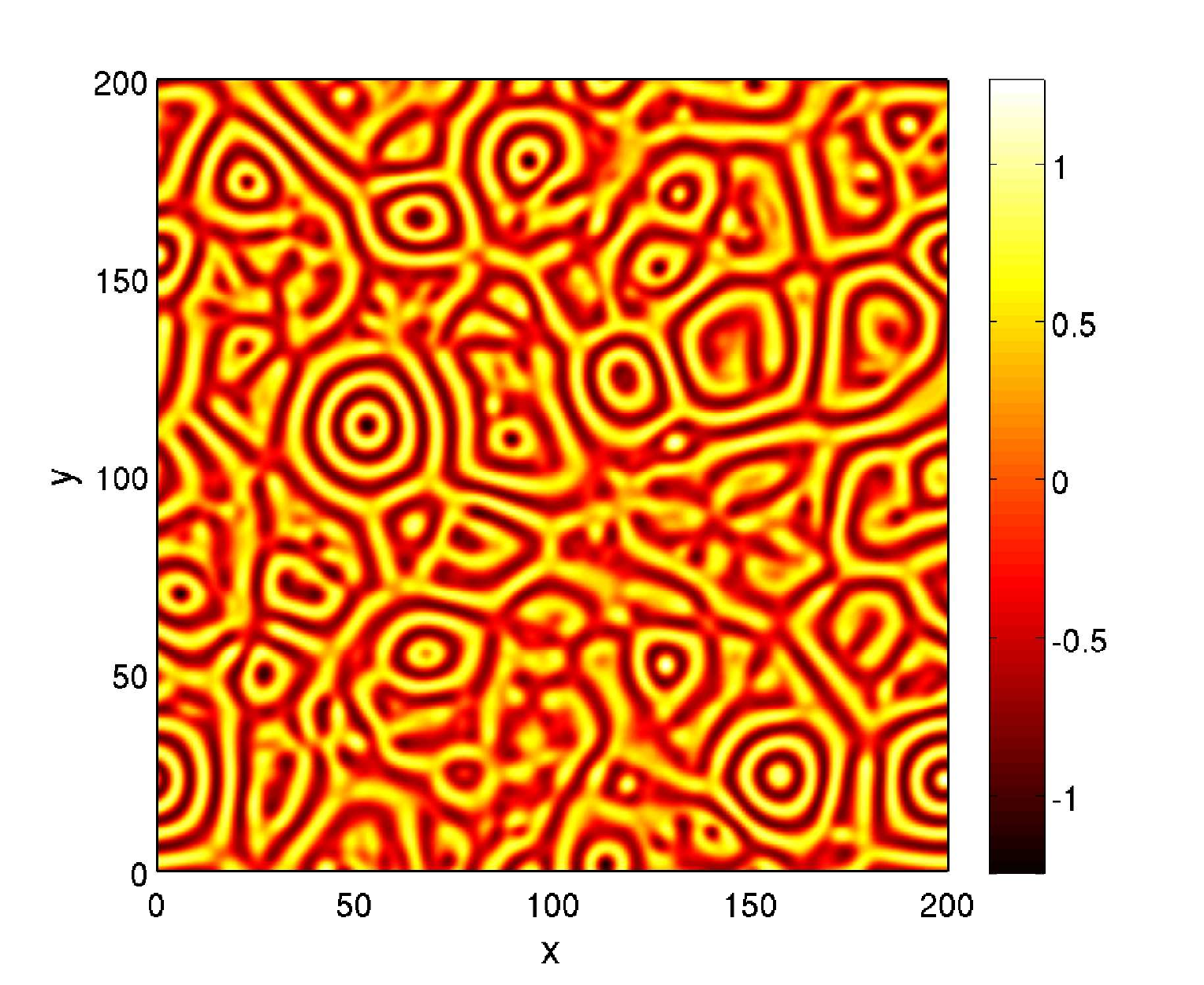}
    \end{subfigure}
    \begin{subfigure}{0.45\textwidth}
        \centering
        \includegraphics[width=0.9\linewidth]{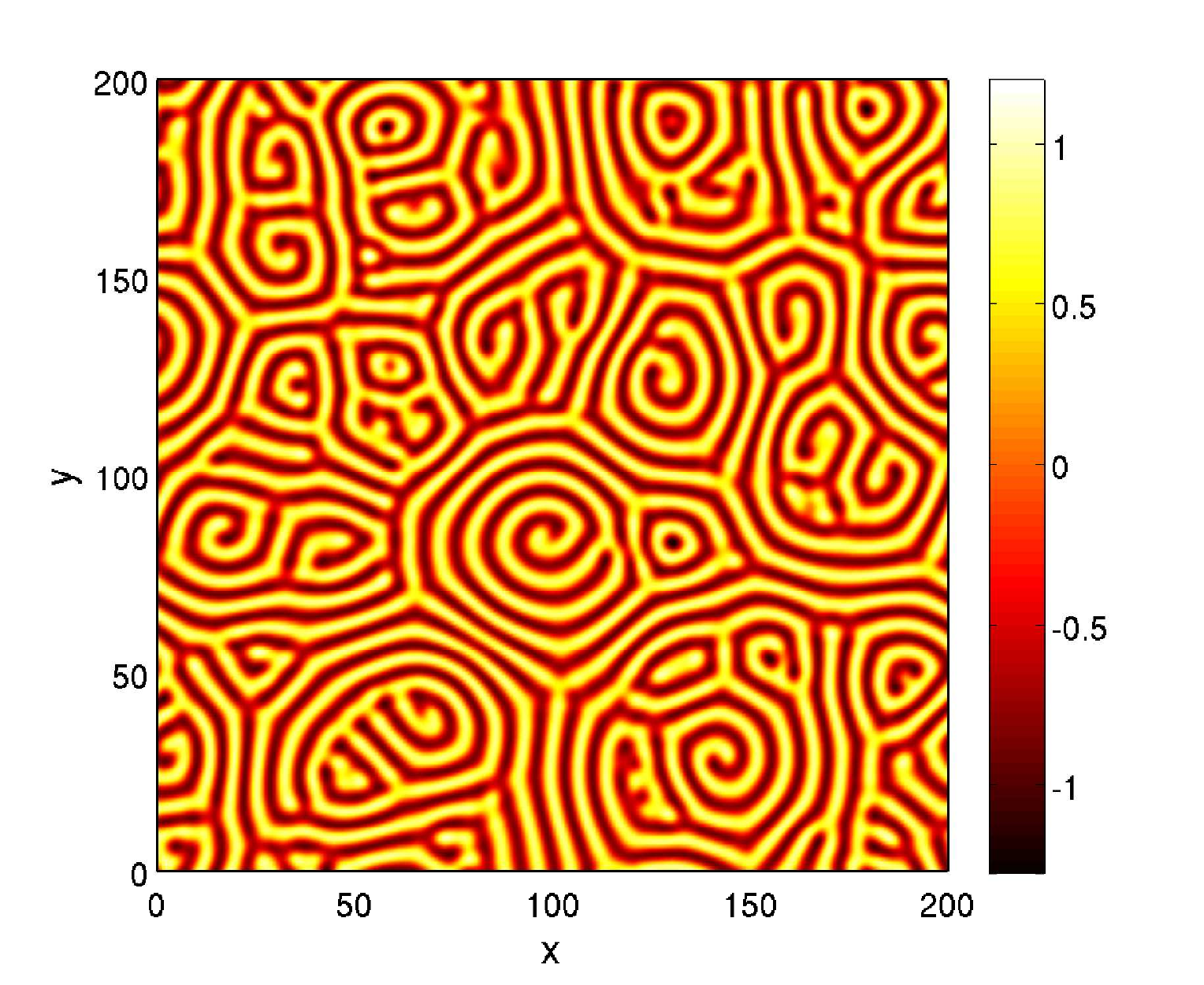}
    \end{subfigure}
    \caption{Patterns of order parameter field $\psi$ obtained from the generalized Swift-Hohenberg model for $\epsilon= 0.7$, $\sigma = 2$, and $c^2 = 0$, at time $t = 2 \times 10^3$. Left: $g_m = 50$, showing a chaotic state without the emergence of spirals. Right: $g_m = 5$, showing spiral defect chaos.}
    \label{fig:c0}
\end{figure}

These results indicate three distinct regimes which can be correlated with the qualitative nature of the system dynamics obtained from the generalized Swift-Hohenberg model. (i) For very small $c^2$ ($\lesssim 0.1$) at $\sigma = 2$, the observed defect patterns are chaotic but without any observable spirals, other than some transient target defects (similarly for $\sigma \lesssim 0.25$ at $c^2 = 1$). As seen in the left panel of Fig. \ref{fig:hydro}, when $c^2 \leq 0.1$ the dynamics are mainly driven by advection and the diffusive dynamics contribution from the RHS of Eq. (\ref{eq:sh}) to $\partial_t\psi$ becomes very small. At $c^2 = 0$ we still observe a few transient targets. (ii) At the other extreme with large $c^2$, the calculations in Ref. \cite{karimi2011exploring} showed that the leading-order (and positive) Lyapunov exponent of the flow approaches zero, indicating very weak or even non chaotic state. The patterns are dominated by slowly coarsening, large target and spiral defects, mixing with small spirals or targets. Similar results can be obtained for large enough $\sigma$ at $c^2 = 1$ in the right panel of Fig. \ref{fig:hydro}. In this regime, diffusive roll unwinding mainly determine spiral rotation, as can be seen in Fig. \ref{fig:hydro}. (iii) In the intermediate parameter range (e.g., around $0.1 < c^2 < 5$ for $\sigma  = 1 \sim 2$ or in the mid-range values of the $\sigma$ dependence at $c^2=1$) spiral defect chaos is observed in the numerical solutions. In this range the contributions from advection and diffusive relaxation are comparable. In particular, both contributions are nearly the same around $c^2 = 1$ and $\sigma = 2$, the parameter values that correspond to the experiment of Ref. \cite{morris1993spiral}, and to the previous study of the Boussinesq equations \cite{paul2003pattern,chiam:2003,re:decker94,karimi2011exploring}. These results also help explain why spiral defect chaos was not observed when $c^2 = 0$ and $g_m = 50$, but did appear by reducing the latter to $g_m = 5$, as shown in Fig. \ref{fig:c0}. That is, reducing $g_m$ would roughly translate into moving the curves shown in Fig. \ref{fig:hydro} to the left. In summary, the results suggest a correlation between the existence of spiral defect chaos and the relative balance between advection and order parameter diffusion.

Characteristic values of the spiral rotation rate $\omega$ can be obtained from the numerical solutions. We estimate  $\omega \sim \mathcal{O}(10^{-1})$ in dimensionless units, with its maximum value close to $1$. This is consistent with the values of $\alpha$ used in the fits of Fig. \ref{fig:vphi}.

\subsection{Comparison with spirals obtained from the Boussinesq equations}
\label{sec:bsq2}

We have explored spiral defect chaos in the Boussinesq model only for no-slip boundary conditions. From the numerical results, as in Fig. \ref{fig:bsqtemp}, we observed that the size of the spirals in the range of parameters where chaos exists is fairly small, as is the case in experiments. Therefore, we cannot examine the asymptotic decay of $v_{\varphi} \sim 1/r$ as has been predicted for large $r$, nor can we conclusively obtain the spatial dependence of long-range flows at small damping as argued above. We do present, however, results for a large, single rotating spiral (see Sec. \ref{sec:bsq}), and evidence that the azimuthal velocity field obtained agrees, without any adjustable parameters, with $v_{\varphi}$ obtained from direct integration of the generalized Swift-Hohenberg equation and in the regime of spiral defect chaos. We therefore expect that the asymptotic behavior of Fig. \ref{fig:vphi} at small damping would carry over to the full Boussinesq model.

\begin{figure}[htp]
    \centering
	\begin{subfigure}[b]{0.37\textwidth}
	\includegraphics[width=\textwidth]{temp_gspiral2.pdf}
	\end{subfigure}
	\begin{subfigure}[b]{0.3\textwidth}
    \includegraphics[width=\textwidth]{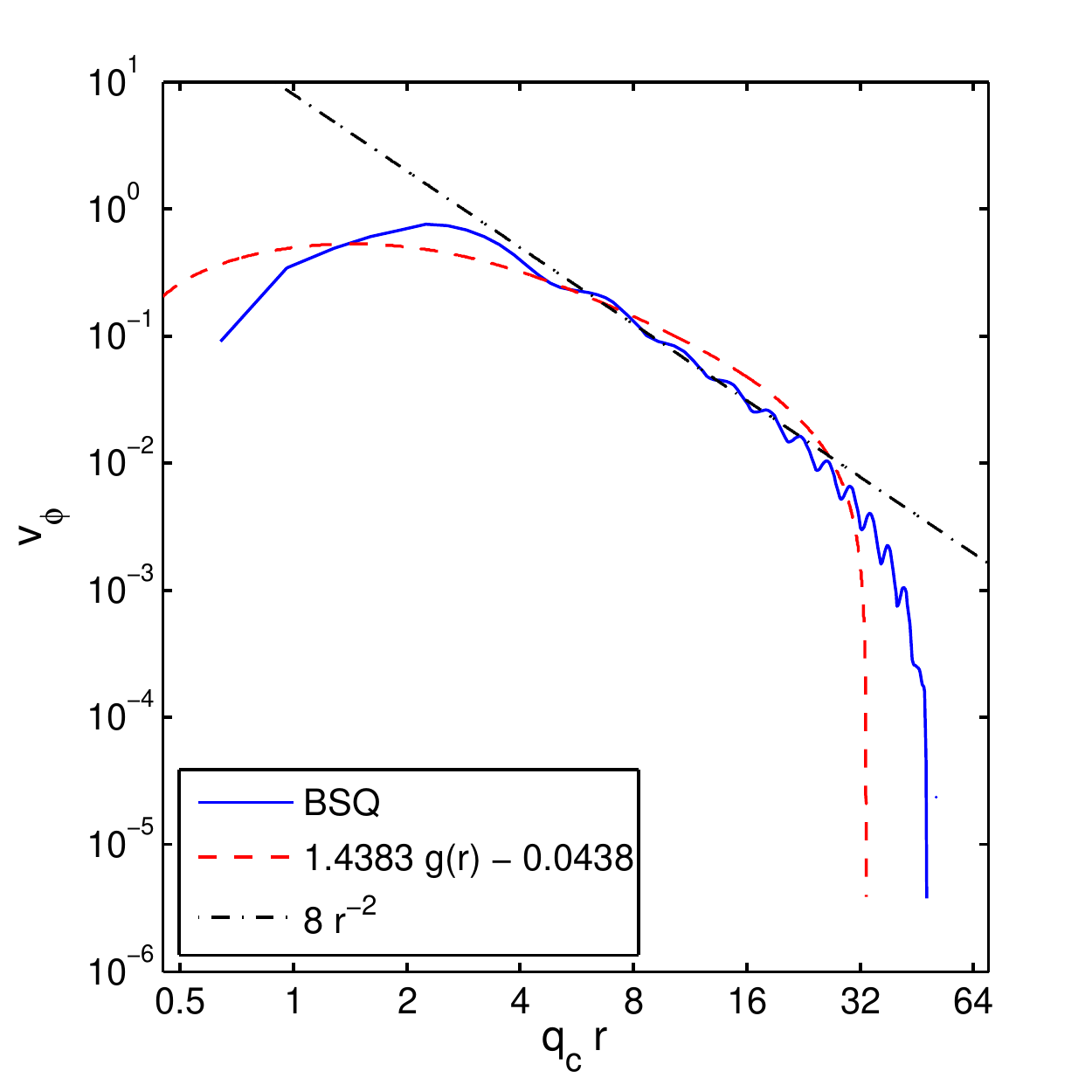}
    \end{subfigure}
    \begin{subfigure}[b]{0.3\textwidth}
    \includegraphics[width=\textwidth]{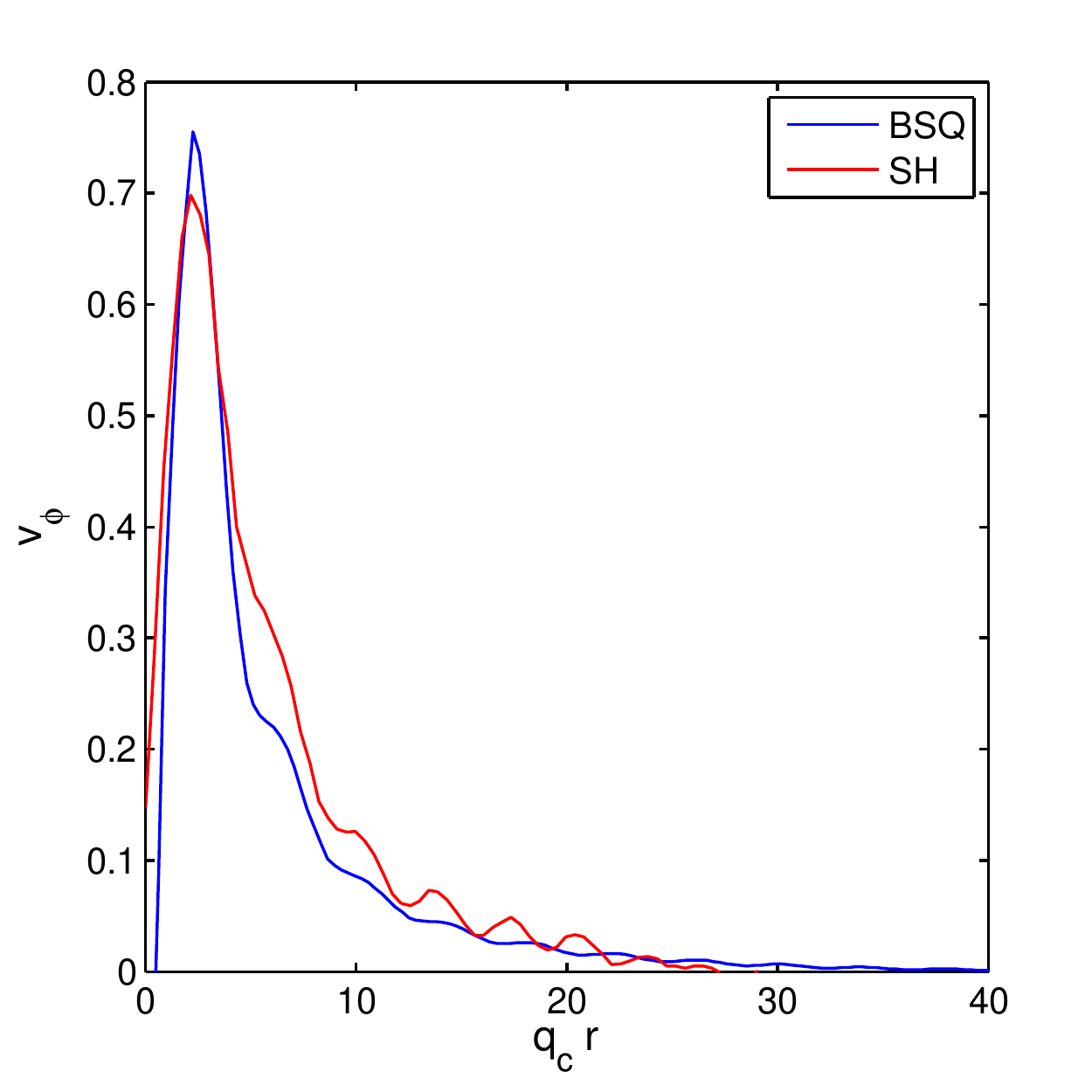}
    \end{subfigure}
    \caption{Left: Temperature field at the midplane for a cylindrical convection cell obtained by time integration of the Boussinesq (BSQ) equations where $q_c = 3.1165$, $\Gamma=40$, $\epsilon=0.4637$, and $\text{Pr}=1$.  Middle: Rescaled azimuthal velocity $v_\varphi$ from the Boussinesq model with $r=0$ at the spiral core (solid line). The straight dashed line illustrates the power law fit $v_{\varphi} \sim r^{-2}$. The red dashed line is the result of Eq. (\ref{eq:bessel}), using $c^2=1$, $\sigma = 2$, $\alpha = 1.4383$, $\beta = -0.04328$, and $r_b = 35 q_c$. Right: $v_{\varphi}$ as a function of $q_{c}r$ obtained from the generalized Swift-Hohenberg model (SH) with $c^2 = 1$, $\sigma = 2$, $\epsilon = 0.7$, and $g_m = 50$, for the spiral located at $(104,56)$ in Fig. \ref{fig:same}, as compared to that of the Boussinesq equations. The $x$ axis is scaled with $q_{c} = 3.1165$ for the Boussinesq result, and $q_{c} = q_{0}$ for the Swift-Hohenberg model. No adjustable parameters have been used.}
	\label{fig:gspiral}
\end{figure}

A configuration comprising a single rotating spiral as given by the Boussinesq model with Prandtl number $\text{Pr} = 1$ is shown in Fig. \ref{fig:gspiral}. It has been obtained by adding a lateral forcing thermal boundary term in a cylindrical cell (a hot sidewall) while setting no-slip boundary conditions at all material surfaces, as described in Sec. \ref{sec:bsq}. The temperature field at mid cell of a slowly rotating spiral and an accompanying dislocation is shown in Fig. \ref{fig:gspiral}. Lengths are made dimensionless by the cell thickness, so that the critical wavenumber is $q_c = 3.1165$, which can be obtained from the marginal stability problem at the critical Rayleigh number. The size of the spiral in Fig.~\ref{fig:gspiral} is 14 wavelengths, before reaching the dislocation. According to the derivation by Manneville \cite{manneville1984modelisation}, given no-slip boundary conditions and a cell with dimensionless thickness $h = 1$, a Galerkin expansion of the flow indicates that the mean flow becomes Poiseuille-like at lowest order. By averaging the governing equations over the height, a vorticity equation analogous to Eq.~(\ref{eq:vort}) can be obtained, with a damping coefficient $c^2 = 10$ corresponding to $q_c = 3.1165$. More details are given in Appendix \ref{sec:ap}, including how the length, time and various parameters are mapped from the Boussinesq model with no-slip conditions and $q_c = 3.1165$, to the generalized Swift-Hohenberg model, Eqs. (\ref{eq:sh}) and (\ref{eq:vort}) with $q_0 = 1$. The value of $c^{2}$ is further rescaled by $1/q_c^2$, so that $c^2 \approx 1$. Since the length scales as $1/q_c$ and for $\text{Pr} = 1$ the time scale is $2.05/q_c^2$, the Boussinesq velocity is rescaled by $2.05/q_c$ to agree with our dimensionless units. Finally, based on the Prandtl and Rayleigh numbers of the Boussinesq solution, we have $\sigma = 2$, $\epsilon = 0.7$, and $g_m = 50$, with scaling also given in Appendix \ref{sec:ap}.

Figure \ref{fig:gspiral} (middle) shows the rescaled azimuthal velocity $v_\varphi$ computed at the mid plane of the cell as a function of $q_{c}r$, so that we can compare it directly with the result from the generalized Swift-Hohenberg model. The coordinate origin has been placed at the spiral's core. Following an initial rise from zero at the core, the velocity appears to decay with distance as $r^{-2}$ between $q_{c}r=3$ and $q_{c}r = 20$, until it quickly decays to zero at about $q_{c} r = 46$. This decay is faster than the expected asymptotic behavior of $1/r$, although we must note that for the time shown, there still is a difference between the rotation velocities of the core and the dislocation given in the simulation.

Figure \ref{fig:gspiral} (right) compares the azimuthal velocity of the large spiral obtained from the Boussinesq model with that of a rotating spiral in the fully chaotic regime given by the generalized Swift-Hohenberg model. We have mapped the physical values of the parameters in the Boussinesq model to the parameters of the generalized Swift-Hohenberg model, as explained in Appendix A. Therefore there are no adjustable parameters. The Swift-Hohenberg result was obtained by computing the adiabatic flow (by eliminating the time derivative in Eq. (\ref{eq:vort})) associated with $\psi$ in Fig. \ref{fig:psi}, for $c^2 = 1$, $\sigma = 2$, $\epsilon = 0.7$, and $g_m = 50$, and the spiral located at the coordinate $(104,56)$. The azimuthal velocity obtained from the Boussinesq model agrees quantitatively with the result of the generalized Swift-Hohenberg model, even when comparing an isolated spiral in the former to one in the chaotic regime in the latter. This lends credence to the observation that the asymptotic calculation of Sec. \ref{sec:vel}, which is based on a rigidly rotating spiral, is a good approximation to the flow induced by a spiral in the chaotic state, albeit in the range of moderate to large values of the damping parameter $c^{2}$. For the value of $c^{2} = 1$ used in the comparison, the azimuthal velocity within a rotating spiral depends only weakly on whether the spiral is isolated or surrounded by other spirals. 

\begin{figure}[htp]
	\centering
    \begin{subfigure}[b]{0.35\textwidth}
    \includegraphics[width=\textwidth]{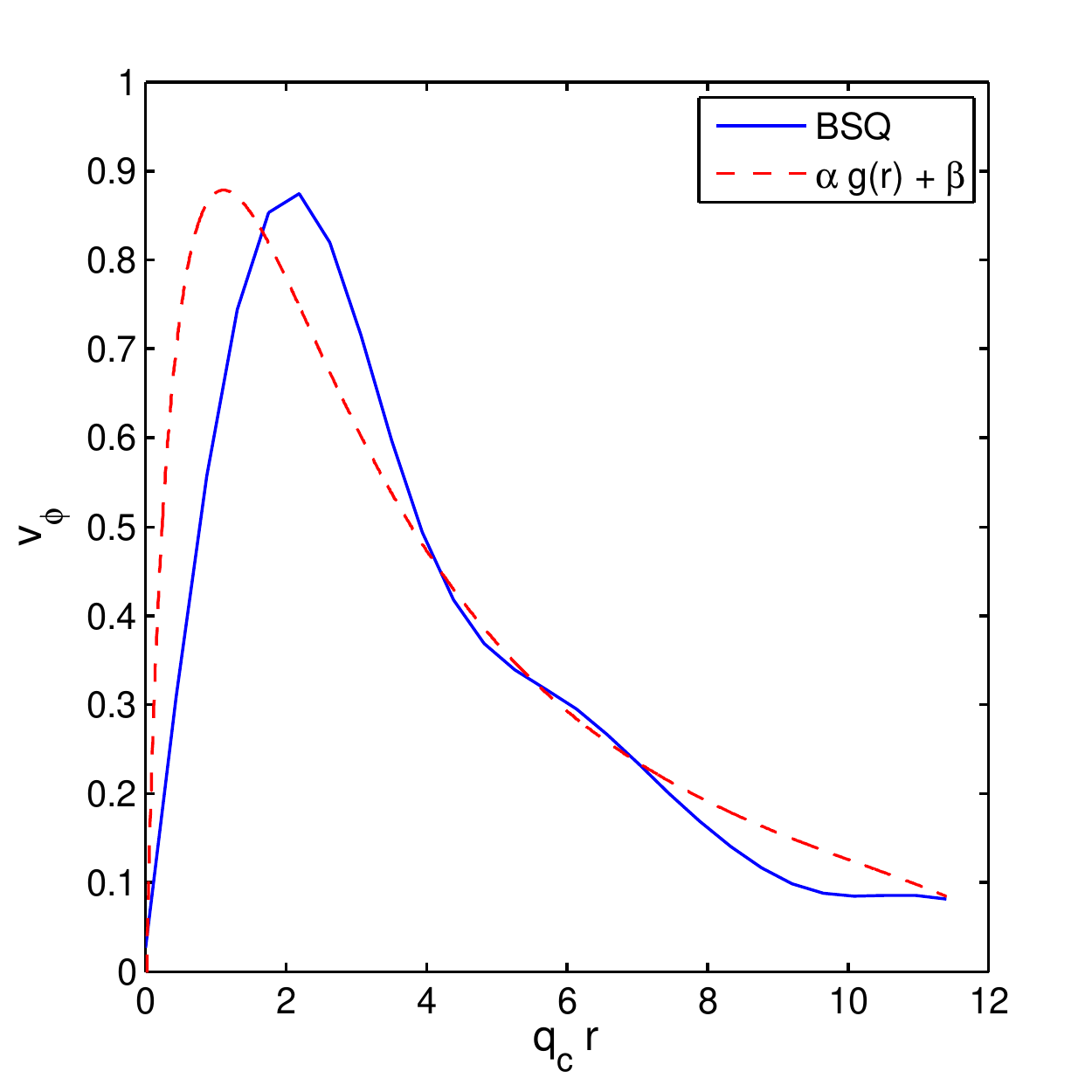}
    \end{subfigure}
    %\hspace{3mm}
    \begin{subfigure}[b]{0.35\textwidth}
    \includegraphics[width=\textwidth]{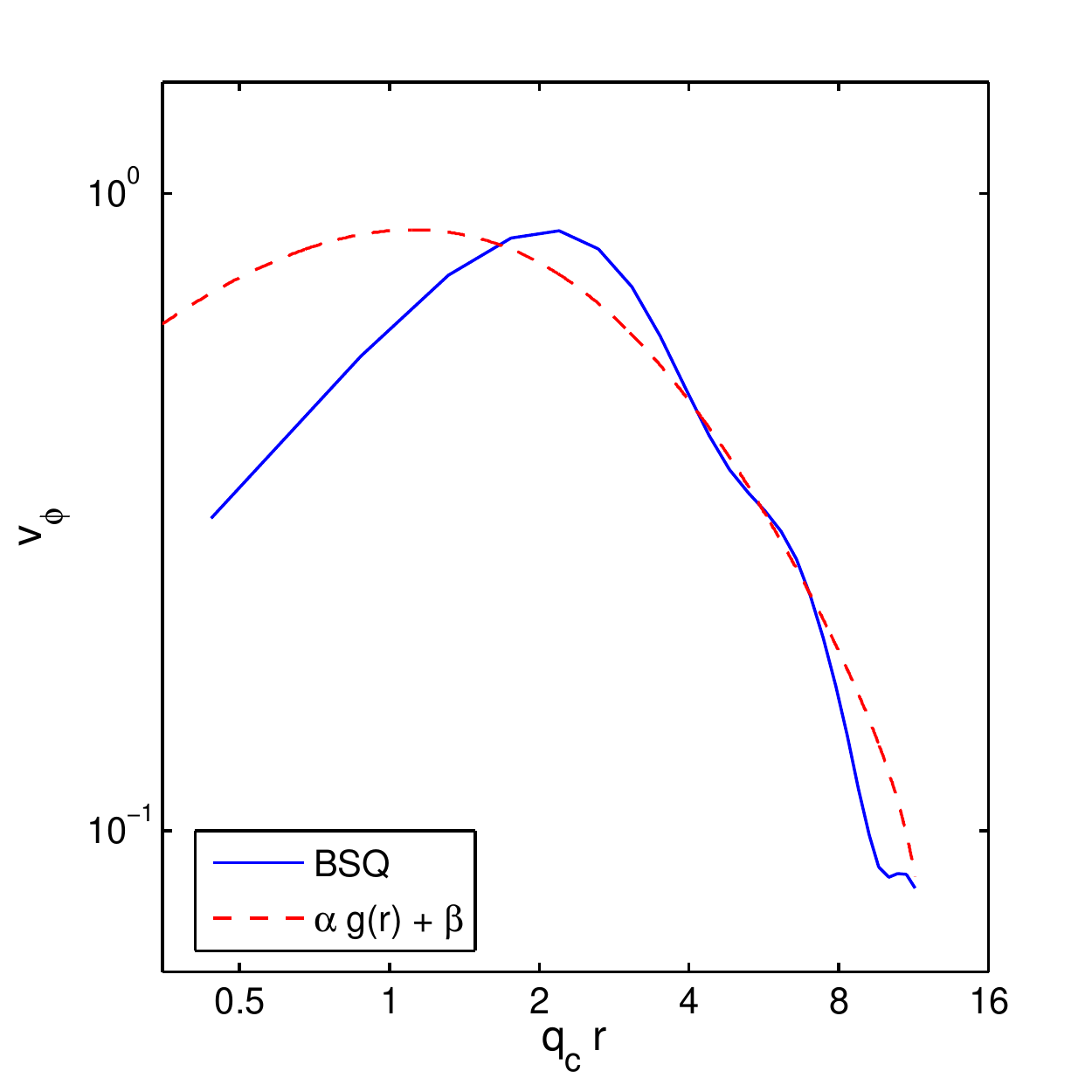}
    \end{subfigure}
    \caption{Rescaled azimuthal velocity for the enlarged spiral in Fig. \ref{fig:bsqtemp} obtained from the Boussinesq model (BSQ), with $r=0$ at its core. The corresponding rescaled parameters are $\sigma = 2$, $c^2 = 1$, and $\epsilon = 0.7$. Left: Comparison of numerical results with analytic predictions. Right: The same results using a logarithmic scale. Parameters of $c^2=1$, $\alpha = 5/\sigma$, $\beta = -0.1$ and $r_b = 4.5$ are used for the analytic curve.}
	\label{fig:bsqvphi}
\end{figure}

We finally analyze the azimuthal velocity in the chaotic regime of the Boussinesq model, with results presented in Fig. \ref{fig:bsqvphi}. We show the azimuthal velocity of the spiral of Fig. \ref{fig:bsqtemp}, with $r=0$ centered at the core of the spiral. This result is compared with our analytic prediction of Eq. (\ref{eq:bessel}), with $c^2 = 1$, $g_m = 50$, $\sigma = 2$, and $\epsilon = 0.7$. Instead of fitting both $\alpha$ and $\beta$ as previously discussed in Sec. \ref{sec:gsh2}, we set $\alpha = 5/\sigma$ (the same value used in Fig. \ref{fig:vphi}). Since $\alpha = m\omega\epsilon g_m/6q_0^2\sigma$, all the parameters used in Sec. \ref{sec:gsh2} are the appropriate ones for the Boussinesq model with no-slip boundary conditions as described in Sec. \ref{sec:bsq}. In both cases we obtain one-armed spirals ($m = \pm 1$). Here we use $r_b = 4.5$ (the approximate size of the spiral) and $\beta = -0.1$. Note that due to the small size of the spiral, $v_\varphi$ does not reach zero at the edge of the spiral, as this is a small defect constrained by other features of the disordered pattern. Even under these circumstances, there is good agreement between the result of the Boussinesq model and the analytic solution, when using the same parameters estimated from the data fit with the generalized Swift-Hohenberg model.

\section{Conclusions}

An irrotational azimuthal body force proportional to $\hat{\bm{\varphi}}/r$ in the generalized Swift-Hohenberg model induces an azimuthal velocity $v_{\varphi}$ for a configuration of a rigidly rotating spirals. At zero damping (free-slip), this force leads to long ranged flows proportional to $-r\ln(r/r_{b})$, where $r_{b}$ needs to be determined independently. For the more realistic case of no-slip boundary conditions, the azimuthal velocity would be expected to decay as $1/r$ instead at a scale $r \gg 1$ when $r_{b} \rightarrow \infty$. For realistic spiral sizes in the regime of spiral defect chaos, $q_{0} r_{b}$ is of order one, and this asymptotic regime is never reached. Instead, the velocity flow depends strongly on the value of $r_{b}$, which in turn depends on the characteristic separation between neighboring spirals.

This dependence of the azimuthal velocity has been compared with direct numerical solution in a chaotic state both for the 3D Boussinesq equations and for the 2D generalized Swift-Hohenberg equations. For free-slip boundary conditions in the latter, the velocity behaves as $-r\ln (r/r_b)$, which is long-ranged and necessarily crosses zero between neighboring spirals of the same topological charge, while for neighboring spirals of opposite charge the velocity interacts constructively. When no-slip conditions are considered (with a finite damping parameter), the velocity profile qualitatively changes. This agrees with our predictions that the velocity decay is governed by a combination of modified Bessel functions. When damping is sufficiently high, we confirm the earlier suggestions that the flow within a spiral is largely independent of the background in which it is immersed. For moderate damping the flow within a spiral is a function of the spiral's size, and hence of the distance to neighboring spirals and their topological charges. This observation is consistent with an earlier suggestion of spiral defect chaos as a form of invasive chaos \cite{cross1996theoretical}, except that hydrodynamic flows also play a role.

For the 2D generalized Swift-Hohenberg model, we identify two contributions to the spiral rotating dynamics: Mean flow advection and diffusive dynamics with wavevector frustration and roll unwinding. We have performed a series of calculations by varying the damping coefficient and the Prandtl number to identify three distinct regimes: Chaos without spiral patterns, diffusive pattern dynamics with extremely weak or no chaos, and spiral defect chaos. The latter appears in the range in which order parameter advection and diffusive relaxation are of similar magnitude. In particular, we find that both contributions are approximately the same at a damping coefficient about $c^2 = 1$, and a rescaled Prandtl number of $\sigma = 2$ (with $\textrm{Pr}=1$), which correspond to the experimental conditions for convection in CO$_2$ gas.

The 3D Boussinesq equations have been integrated in a rectangular geometry with periodic sidewalls and no-slip conditions at top and bottom surfaces. By analyzing the flow field around a small spiral in the chaotic state, we found that the analytic result based on a rigidly rotating spiral agrees reasonably well with the Boussinesq azimuthal velocity, and that the remaining fit coefficients are consistent with those used with the generalized Swift-Hohenberg results under corresponding values of the physical parameters. Finally, we obtained a large spiral using the Boussinesq model in a cylindrical configuration, and analyzed the azimuthal flow around the core of the spiral. The azimuthal velocity agrees with the generalized Swift-Hohenberg result without any adjustable parameters. We conclude that for large values of the damping parameter, the flow field induced by a rotating spiral is the same regardless of whether it is isolated or surrounded by other spirals. When the damping parameter is reduced, the flow field depends on the distance to neighboring spirals and the relative sign of their topological charge, therefore providing a means for their interaction.

\section*{Acknowledgments}

This research has been supported by the Minnesota Supercomputing Institute, and by the Extreme Science and Engineering Discovery Environment (XSEDE) \cite{towns2014xsede} which is supported by the National Science Foundation under grant number ACI 1548562. EV acknowledges a Doctoral Dissertation Fellowship from the University of Minnesota, and support from the Aerospace Engineering and Mechanics department. The research of JV is supported by the National Science Foundation under Grant No. DMR-1838977. MP and SM acknowledge support for computing resources from the Advanced Research Computing center at Virginia Tech. Z-FH acknowledges support from the National Science Foundation under Grant No. DMR-1609625.

\appendix
\section{Parameter values of the generalized Swift-Hohenberg model}
\label{sec:ap}

A generalized Swift-Hohenberg model that includes advection by the solenoidal mean flow velocity $\mathbf{v}$ has been derived by Manneville \cite{manneville1983two,manneville1984modelisation} from the Boussinesq equations, which has the following form
\begin{eqnarray}
&&\tau_0 \left ( \partial \psi / \partial t + {\bm v} \cdot {\bm
    \nabla} \psi \right )
= \left [ \epsilon - \frac{\xi_0^2}{4q_c^2} \left ( \nabla^2 + q_0^2
\right )^2 \right ] \psi 
-g(\text{Pr}) N[\psi], \label{eq:Manneville}\\
&&\left [ \partial / \partial t - \text{Pr} (\nabla^2 - c^2) \right ] \nabla^2 \zeta
= g_{q_c} \left [ {\bm \nabla}(\nabla^2 \psi) \times {\bm \nabla} \psi
\right ] \cdot \hat{\bm z}, \label{eq:mean_flow_M}
\end{eqnarray}
where
\begin{equation}
{\bm v} = {\bm \nabla} \times (\zeta \hat{\bm z})
= \left ( \partial_y \zeta, -\partial_x \zeta \right ).
\label{eq:U}
\end{equation}
In Manneville's model, the nonlinearity $N[\psi]$ has the form $N[\psi] = |\bm{\nabla}\psi|^2\psi+q_c^2\psi^3$. However, the threshold expansion of the Boussinesq equations leads to the cubic sum of Fourier modes, with no counterpart in real space \cite{cross1980derivation}. Therefore, there is no systematic way for which $N[\psi]$ can be derived for a real-space expression, and its form depends on boundary conditions and arbitrarities of expansions \cite{greenside1985stability}. In this work we use the simplest form $N[\psi] = q_c^2\psi^3$. The model parameters depend on boundary conditions for the top and bottom of the convection cell. In the case of no-slip (rigid) boundary conditions, where $c^2 > 0$ accounting for hard-mode oscillatory instabilities, these parameters are given by \cite{manneville1984modelisation}
\begin{eqnarray}
&\epsilon = (\text{Ra}-\text{Ra}_c)/\text{Ra}_c, \qquad \text{Ra}_c=1750 ~\textrm{(exact value: 1708)}, \qquad c^2=10& \nonumber\\
&q_c=3.1165 ~(\sim \text{exact value}), \qquad \xi_0^2=0.1497
~{\rm (exact ~value: 0.148)},& \nonumber\\
&\tau_0=(1.9425+{\textrm{Pr}}^{-1})/38.2927 ~\text{[exact value:}
~(1.9544+{\text{Pr}}^{-1})/38.4429],& \nonumber\\
&g_{q_c}=2/(21q_c), \qquad g(\text{Pr})=\alpha_0 + \beta_0/\text{Pr} +
\gamma_0/{\textrm{Pr}}^2,& \label{eq:values}
\end{eqnarray}
where $\alpha_0, \beta_0, \gamma$ are some unknown expansion coefficients. Note that the above parameters such as $\text{R}_c$, $q_c$, $\xi_0$, and $\tau_0$ were derived from the  Galerkin expansion by Manneville \cite{manneville1983two,manneville1984modelisation} and well agree with the known exact values; also the length scale used above should be the vertical thickness $d$, and hence the dimensional $q_c \rightarrow q_cd$ and $c^2 \rightarrow c^2 d^2$ after rescaling.

In our simulations (and most other research), the dimensionless model equations are used. Setting a length scale $1/q_c$, a time scale $4\tau_0/ (\xi_0^2 q_c^2)$, the rescaled variables $\psi' = \psi\sqrt{4g(Pr)/\xi_0^2}$, $\zeta'=\zeta (4\tau_0/\xi_0^2)$, as well as
\begin{equation}
\epsilon'=\epsilon \frac{4}{\xi_0^2q_c^2} = \frac{\text{Ra}-\text{Ra}_c}{\text{Ra}_c}
\left ( \frac{4}{\xi_0^2q_c^2} \right ), \qquad c'^2=c^2/q_c^2,
\end{equation}
and omitting all the primes, the generalized Swift-Hohenberg model equations (\ref{eq:Manneville}) and (\ref{eq:mean_flow_M}) can be rescaled as
\begin{eqnarray}
&\partial \psi / \partial t + {\bm v} \cdot {\bm \nabla} \psi
= \left [ \epsilon - \left ( \nabla^2 + q_0^2 \right )^2 \right ] \psi
- N[\psi],& \label{eq:model}\\
&\left [ \partial / \partial t - \sigma (\nabla^2 - c^2) \right ] \nabla^2 \zeta
= g_m \left [ {\bm \nabla}(\nabla^2 \psi) \times {\bm \nabla} \psi
\right ] \cdot \hat{\bm z},& \label{eq:mean_flow}
\end{eqnarray}
as used in our study. Here
\begin{equation}
q_0^2=1, \qquad \sigma=\frac{4\tau_0}{\xi_0^2} \text{Pr}, \qquad 
g_m=\frac{4\tau_0^2}{g(\text{Pr})\xi_0^2}g_{q_c}.
\label{eq:para}
\end{equation}
From the values given in Eq. (\ref{eq:values}), the parameters in the above equations (\ref{eq:model}) and (\ref{eq:mean_flow}) can be estimated as
\begin{eqnarray}
&\epsilon=2.7511 (\text{Ra}-\text{Ra}_c)/\text{Ra}_c, \qquad c^2 = 10/q_c^2 = 1.03,&\nonumber\\
&\sigma=0.6978 (1+1.9425 \text{Pr}), \qquad 
g_m=1.7868 \times 10^{-4} (1.9425+{\text{Pr}}^{-1})^2/g(\text{Pr}).&
\end{eqnarray}
If the Prandtl number $\text{Pr}=1$ as set in experiments and the simulations of the Boussinesq model, we have $\sigma \simeq 2$. Also if choosing $g(\text{Pr})=3.0941\times 10^{-5}$, we get $g_m=50$ as used in our calculations. In many calculations $\epsilon$ is set as $0.7$, which corresponds to $(\text{Ra}-\text{Ra}_c)/\text{Ra}_c=0.2544$. This choice started from the first theoretical paper of spiral defect chaos \cite{xi1993spiral}, based on the experimental results showing the onset of spiral chaos at $(\text{Ra}-\text{Ra}_c)/\text{Ra}_c \geq 0.25$ for $\text{Pr}=1$ and in systems of large enough aspect ratio \cite{morris1993spiral,re:liu96}. 
The value of $c^2$ [$=10$ (unscaled) or equivalently $\simeq 1$ after rescaling] comes from the approximation process based on no-slip boundary condition \cite{manneville1984modelisation}; after rescaling it is independent of the Prandtl number or the length scale chosen. Values of $g(\text{Pr})$, $g_{q_c}$, and hence $g_m$ also depend on the approximation of expansion (or the averaging over vertical thickness). As pointed out by Manneville \cite{manneville1984modelisation}, their values would be different if using a different averaging procedure (e.g., the unscaled $c^2$ would change from $10$ to $12$, and $g_{q_c}$ from $2/21q_c$ to $4/35q_c$).

We note that in most of previous studies using the generalized Swift-Hohenberg equations, usually $\sigma$ is set as $1$ which actually corresponds to $\text{Pr} \simeq 0.22$; also $c^2=2$ was first chosen in Ref. \cite{xi1993spiral} and then followed in almost all the later work except for Ref. \cite{karimi2011exploring} which explored a range of possible $c^2$ values.

\bibliography{sdc}

\begin{thebibliography}{50}
\bibitem [{re:()}]{re:misc1}%
  \BibitemOpen
  \href@noop {} {}\bibinfo {note} {Note that the flow itself is rotational,
  i.e., $\bm{\nabla} \times \mathbf{v}_{1} \neq 0$. While the force component
  $\bm{f}_{\varphi} = -(\rho^2m\omega/r)\bm{\hat{\varphi}}$ is irrotational,
  its curl is nonzero, since $\bm{\nabla} \times (\bm{f}_{\varphi}) \propto
  \delta(0) \bm{\hat{z}}$. Hence, it leads to a nonzero vorticity, and
  consequently to the rotational $\mathbf{v}_{1}$.}\BibitemShut {Stop}%
\end{thebibliography}

\end{document}